\documentclass[12pt,epsf,epsfig,aps,floatfix,nofootinbib,superscriptaddress]{revtex4}
\usepackage{graphics}
\usepackage{graphicx}
\usepackage[english]{babel}
\usepackage{amsmath}

\newcommand{\nc}{\newcommand}
\newcommand{\be}{\begin{equation}}
\newcommand{\ee}{\end{equation}}
\newcommand{\beq}{\begin{eqnarray}}
\newcommand{\eeq}{\end{eqnarray}}
\nc{\barray}{\begin{eqnarray}}
\nc{\earray}{\end{eqnarray}}
\nc{\barrayn}{\begin{eqnarray*}}
\nc{\earrayn}{\end{eqnarray*}}
\nc{\bcenter}{\begin{center}}
\nc{\ecenter}{\end{center}}
\nc{\ket}[1]{| #1 \rangle}
\nc{\bra}[1]{\langle #1 |}
\nc{\0}{\ket{0}}
\nc{\mc}{\mathcal}
\nc{\er}[1]{(\ref{eq:#1})}
\nc{\onehalf}{\frac{1}{2}}
\nc{\partialbar}{\bar{\partial}}
\nc{\psit}{\widetilde{\psi}}
\nc{\Tr}{\mbox{Tr}}
\nc{\ev}{\;\mathrm{eV}}
\nc{\mev}{\;\mathrm{MeV}}
\nc{\gev}{\;\mathrm{GeV}}
\nc{\tev}{\;\mathrm{TeV}}

\newcommand{\sigv}{\langle \sigma v \rangle}
\newcommand{\mpl}{M_\text{pl}}

\def\chii0{\chi_i^0}
\def\chij0{\chi_j^0}

\newcommand{\gsim}{\lower.7ex\hbox{$\;\stackrel{\textstyle>}{\sim}\;$}}
\newcommand{\lsim}{\lower.7ex\hbox{$\;\stackrel{\textstyle<}{\sim}\;$}}

\newcommand{\eq}[1]{Eq.~(\ref{#1})}

\begin{document}

\title{Asymmetric Dark Matter: Theories, Signatures, and Constraints}
\author{Kathryn M. Zurek}
\affiliation{Michigan Center for Theoretical Physics, Department of Physics, University of Michigan, Ann Arbor, Michigan 48109 USA}
\email{kzurek@umich.edu}

\begin{abstract}

We review theories of Asymmetric Dark Matter (ADM), their cosmological implications and detection.  While there are many models of ADM in the literature, our review of existing models will center on highlighting the few common features and important mechanisms for generation and transfer of the matter-anti-matter asymmetry between dark and visible sectors.  We also survey ADM hidden sectors, the calculation of the relic abundance for ADM, and how the DM asymmetry may be erased at late times through oscillations.  We consider cosmological constraints on ADM from the cosmic microwave background, neutron stars, the Sun, and brown and white dwarves.  Lastly, we review indirect and direct detection methods for ADM, collider signatures, and constraints.   

\end{abstract}

\maketitle

\newpage

\tableofcontents

\newpage

\section{Motivation: What is Asymmetric Dark Matter?}
\label{sec:intro}

The dark matter (DM) and baryon abundances are very close to each other observationally: $\rho_{DM}/\rho_B \approx 5$ \cite{Ade:2013ktc}.  In the standard Weakly Interacting Massive Particle (WIMP) paradigm, however, these quantities are not {\em a priori} related to each other.  The DM density in the WIMP freeze-out paradigm is fixed when the annihilation rate drops below the Hubble expansion \cite{Scherrer:1985zt,Kolb:1990vq}:
\begin{equation}
n(T_{fo}) \langle \sigma_{ann} v \rangle < H(T_{fo}),
\end{equation}
where $T_{fo}$ is the temperature when DM annihilation freezes-out, $n(T_{fo})$ is the DM number density, and $\langle \sigma_{ann} v \rangle$ is a thermally averaged annihilation cross-section.  Thus the macroscopic quantity of the DM number density in the universe today is related to the microscopic quantity of the annihilation cross-section. On the other hand in baryogenesis \cite{Cohen:1993nk,Riotto:1999yt,Morrissey:2012db}, the baryon density is set by CP-violating parameters and out-of-equilibrium dynamics (such as order of the electroweak phase transition) associated with baryon number violating processes.  Since the quantities setting the baryon density and the DM density are unrelated to each other in these scenarios, it seems surprising that the observed energy densities are so close to each other.  While it is possible that this is an accident, or that this ratio is set anthropically, dynamics may also play a role.  The theory of DM may, in fact, tie the DM density to the baryon density.  

The connection between the DM and baryon densities arises naturally when the DM has an asymmetry in the number density of matter over anti-matter similar to baryons.\footnote{In some theories connecting the DM and baryon densities, the DM does not have a matter-anti-matter asymmetry.  Even though the DM is not asymmetric in these cases, we discuss these models in this review where appropriate.}  The DM density is then set by its asymmetry, which can be directly connected to the baryon asymmetry, rather than by its annihilation cross-section.   Thus we have
\begin{equation}
n_X - n_{\bar{X}} \sim n_b - n_{\bar{b}},
\label{ADMrelation}
\end{equation}
where $n_X,~n_{\bar{X}}$ are the DM and anti-DM number densities, and $n_b,~n_{\bar{b}}$ are the baryon and anti-baryon asymmetries.
The asymmetry is approximately one part in $10^{10}$ in comparison to the thermal abundance, since
\begin{equation}
\eta  \equiv \frac{n_B}{n_\gamma} = \frac{n_b - n_{\bar{b}}}{n_\gamma} \approx 6 \times 10^{-10},
\label{eta}
\end{equation}
with the last relation being obtained most precisely from Cosmic Microwave Background (CMB) data \cite{Ade:2013zuv}.  Since $\rho_{DM}/\rho_B \sim 5$,  the relation of Eq.~\ref{ADMrelation} suggests $m_X \sim 5 m_p \simeq 5 \mbox{ GeV}$. The natural asymmetric DM mass may differ from this value by a factor of a few due to the details of the model.\footnote{This natural relationship is broken in two instances.  First, if DM-number violating process creating the DM asymmetry decouples (at a temperature $T_D$) after the DM becomes non-relativistic, in which case there is a Boltzmann suppression in the asymmetry which scales as $e^{-m_X/T_D}$, where $m_X$ is the DM mass.  Thus the DM can be much heavier than 5 GeV.  Second, if the DM and baryon setting mechanism yields very different asymmetries in the visible and dark sectors, the DM may be much heavier or lighter than 5 GeV.  We will review models that realize both cases, with the former occurring most prominently in sphaleron models, and the latter occurring most prominently in decay models.}  Furthermore, since this scale is not far from the weak scale, in some models the DM mass may be related to weak scale dynamics, reducing the question of why the baryon and DM densities are close to each other to the question of why the weak scale is close to the QCD confinement scale.  In other models, the DM mass scale is set by the proton mass scale itself.

The idea that the DM and baryon asymmetries might be related to each other dates almost from the time of the WIMP paradigm itself \cite{Nussinov:1985xr,Gelmini:1986zz}.  The initial motivation for a DM asymmetry was to solve the solar neutrino problem, by accumulating DM that affects heat transport in the Sun, as pointed out by \cite{Spergel:1984re}.  The subsequent development of DM models with an asymmetry focused on electroweak sphalerons to relate the baryon and DM asymmetries \cite{Barr:1990ca,Barr:1991qn,Kaplan:1991ah,Gudnason:2006ug,Gudnason:2006yj}, though such models usually involve electroweak charged DM, and have become highly constrained by both LEP and the LHC.  In other cases decay mechanisms were utilized \cite{Kitano:2004sv,Kitano:2005ge}.  The Asymmetric Dark Matter (ADM) paradigm \cite{Kaplan:2009ag} provided a robust framework to relate the baryon and DM number densities via higher dimension operators; it encompasses many realizations and easily evades all experimental constraints.  With this paradigm as a sound and flexible framework, significant activity and development of ADM models and phenomenology ensued.  This development is the subject of this review. 
More generally, the ADM mechanism\footnote{While the name ``Asymmetric Dark Matter'' was introduced in \cite{Kaplan:2009ag} to describe the higher dimension operator models proposed there, we use the name ``ADM'' in this review to describe all models where the dark matter density is set via its chemical potential.} works as follows.
\begin{enumerate}
\item An asymmetry is created in the visible and/or dark sectors.  The asymmetry may be created via standard baryo- or lepto-genesis, and then communicated to the DM sector; it may be generated in the DM sector and then transferred to the baryons and leptons in the visible sector; or, a baryon and DM asymmetry may be generated simultaneously.
\item The process which communicates the asymmetry between sectors decouples, separately freezing in the asymmetry in the visible and dark sectors.
\item If the dark sector was thermalized in the process of asymmetry generation, the symmetric abundance (which is $10^{10}$ times larger than the asymmetric component, since the cosmological baryon asymmetry is $\eta \approx 6 \times 10^{-10}$) must efficiently annihilate away.  By analogy with the Standard Model (SM) sector, the most efficient way this is done is via annihilation to force carriers.  For example, $e^+ e^-$ annihilates to photons until only the component fixed by the baryon asymmetry remains.  In the presence of light dark forces, a similar process occurs for DM, though other mechanisms (such as higher dimension operators) may also be at work.
\end{enumerate}
These steps will provide a basic framework for understanding and reviewing models of ADM.

There are many paths to creating this cosmological history via a particle physics model.  However most of these paths fall into a few categories, under asymmetry {\em transfer} and {\em generation}.  Transfer mechanisms work hand-in-hand with existing baryo- or lepto-genesis scenarios, taking an existing primordial asymmetry in baryons or leptons and communicating it to the DM sector.   These transfer mechanisms fall, in general terms, into two categories:
\begin{itemize}
\item electroweak sphalerons, and
\item higher dimension and renormalizable interactions.
\end{itemize}  
In addition, ADM models also have important implications for generation of the asymmetry.  These fall into two categories.
\begin{itemize}
\item Simultaneous generation of baryon and DM asymmetries, which we call  {\em cogenesis}.  Cogenesis may occur via modifications to existing lepto- or baryo-genesis scenarios, that incorporate concurrent DM asymmetry generation, such as out-of-equilibrium decay, the Affleck-Dine mechanism \cite{Affleck:1984fy}, or electroweak baryogenesis.
\item Asymmetry generation in the DM sector, which is then communicated via one of the two transfer mechanisms.  We call this mechanism {\em darkogenesis}.  The asymmetry can be generated in the DM sector by models that mimic some of the successful features of existing baryogenesis scenarios, like electroweak baryogenesis \cite{Cohen:1993nk,Morrissey:2012db} or spontaneous baryogenesis \cite{Cohen:1988kt,Cohen:1991iu}.
\end{itemize}
There are many models in the literature that realize these mechanisms.   Our aim in this review is to illuminate the {\em common features and important mechanisms} of ADM models.  We only detail particular models when its purpose is to illustrate a general mechanism, though we have provided the reader a complete (to our knowledge) list of references to ADM models for further study.   We refer the reader to \cite{Davoudiasl:2012uw,Petraki:2013wwa} for other reviews.

The outline of this review is as follows.  In the next two sections we overview the basic mechanisms for transferring (Sec.~\ref{sec:models}) or generating (Sec.~\ref{sec:baryo}) an asymmetry (step 1 above).  We will explain how the decoupling process (step 2 above) happens naturally in each class of models, from electroweak sphalerons to higher dimension operators, to decay models.   Next, in Sec.~\ref{sec:annihilation}, we consider how the thermal symmetric abundance of DM may be removed by annihilation to SM states or to dark states, including dark forces, in the hidden sector (step 3 above).  We review how the standard thermal freeze-out calculation is modified in the presence of an asymmetry.  This leads naturally to a discussion of dark hidden sectors in Sec.~\ref{sec:hidden sectors}, and how the dynamics in the DM sector sets mass scales and densities, giving rise in many cases to highly non-trivial dynamics.  Even if the DM is asymmetric in the early universe, the asymmetry can be washed out late in the history of the universe; we review how this occurs in Sec.~\ref{sec:oscillations}.
 
Having laid out the general framework for ADM we turn to discussing the constraints on ADM.  In Sec.~\ref{sec:cosmoconstraints} we discuss astrophysical objects such as neutron stars and the impact of ADM on the evolution of the Sun.  We then focus on observation of ADM through direct and indirect detection experiments in Sec.~\ref{sec:directindirectdetection}.  Lastly in Sec.~\ref{sec:colliderconstraints} we review collider and flavor constraints, focusing on extended cascade decay chains in supersymmetry, on displaced vertices, and on their connection to flavor physics.

\section{Mechanisms for Sharing a Primordial Dark or Baryon Asymmetry}
\label{sec:models}

Existing classes of ADM utilize two distinct mechanisms for sharing the asymmetry between sectors.  

The first class makes use of baryon ($B$) and lepton ($L$) number violation via non-perturbative field configurations, or sphalerons.  Electroweak sphalerons are well-known to violate $B$ and $L$ but conserve $B-L$.  In sphaleron transfer models, the DM has a global symmetry, $D$, like $B,~L$, which is conserved in the classical Lagrangian.  Electroweak sphalerons may, however, violate $D$ if $D$ has a chiral anomaly under $SU(2)$.  This is exactly analogous to the way electroweak sphalerons violate $B$ and $L$ in the SM.  DM number then freezes-in when the electroweak sphalerons decouple.  Because of the need to extend the electroweak content of the SM, these models are often subject to severe precision electroweak constraints.  Sphaleron models can, however, be extended to include a new gauge group under which $D$, as well as $B$ or $L$, is instead anomalous, alleviating precision electroweak constraints.

The second class naturally evades precision electroweak constraints by making use of higher dimension operators to transfer a primordial chemical potential in either the visible or DM sectors.  The transfer mechanism explicitly relates the $B$ and $L$ asymmetries to the DM asymmetry via the interactions through the higher dimension operator.  When the operator decouples as the universe cools, asymmetries freeze in separately in each sector.  Renormalizable interactions have also been employed, especially when the DM and/or visible sectors are extended to include additional fields.  In this case the interaction decouples as the new heavy particles become integrated out, making a natural connection between models with higher dimension and renormalizable interactions.

We now  review these two classes of models for asymmetry transfer.

\subsection{Sphalerons}
\label{subsec:sphalerons}

The DM particle (which we will call $X$ throughout this review) carries a global Dark number $D$ analogous to baryon and lepton number, $B$ and $L$.  The associated $U(1)_D$ global symmetry may be violated via sphalerons.
We consider first the case that electroweak sphalerons, as {\em e.g.} in technicolor models \cite{Barr:1990ca,Gudnason:2006yj,Gudnason:2006ug,Foadi:2008qv,Ryttov:2008xe,Frandsen:2009mi,Frandsen:2011kt,Lewis:2011zb}, distribute an asymmetry between DM, baryon and lepton number. 
  
By computing the divergence of the baryon, lepton and DM numbers from
\beq
\partial_\mu j^\mu = \frac{N g^2}{64 \pi^2} \epsilon^{\mu \nu \rho \sigma} F_{\mu \nu}^a F_{\rho \sigma}^a,
\eeq
with $N$ $SU(2)$ doublets, one can determine which linear combinations of $B,~L$ and $D$ are violated by the electroweak sphalerons.  For fields carrying $D$, this divergence is, as usual, proportional to a trace (over all fields carrying $SU(2)$ and $D$ which can contribute to the anomaly) of the $SU(2)$ generators:
\beq
\mbox{tr}\left[\tau^a \tau^b D\right]= \frac{1}{2}\delta^{ab}\sum_i D_i,
\eeq 
where $\tau_a,~\tau_b$ are the $SU(2)$ generators, and $D_i$ is the $U(1)_D$ charge of $i$th particle carrying $SU(2)$ in the theory.  If this sum is non-vanishing, $D$ is violated by the electroweak sphalerons.

With $N_g$ generations carrying $B$ and $L$, and $N_X$ DM electroweak doublets, by computing the anomalies of $B$, $L$ and $D$ with $SU(2)$, one finds that the sphaleron carries $B = N_g/2$, $L = N_g/2$ and $D = N_X/2$.  As a result, sphalerons violate $B + L + \frac{N_X}{N_g} D$, while they conserve $I_1 = B - L$ and $I_2 = B - \frac{N_g}{N_X} D$.  For example, an early concrete model \cite{Kaplan:1991ah} introduced a fourth generation, with the quark sector of the fourth generation mixing with the SM quarks, but the leptonic fourth generation having an exactly conserved $D$ symmetry that is enforced by no mixing with the SM sector.  In this case $I_2 = B - 4 D$ is conserved, and as a result
\beq
\Omega_B/\Omega_{DM} = 4 m_p/m_X,
\eeq  
where $m_X$ is the mass of the DM, so that the DM is not much heavier than the proton. Since the DM is part of an $SU(2)$ multiplet, the $Z$ can decay to the DM.  The DM may also scatter through the $Z$ in direct detection experiments, which leads to scattering cross-sections $\sim 10^{-39} \mbox{ cm}^2$.  Direct detection experiments and the invisible $Z$ width thus rule out the simplest version of this model.

\begin{figure}
\begin{center}
\includegraphics[width=0.42\textwidth]{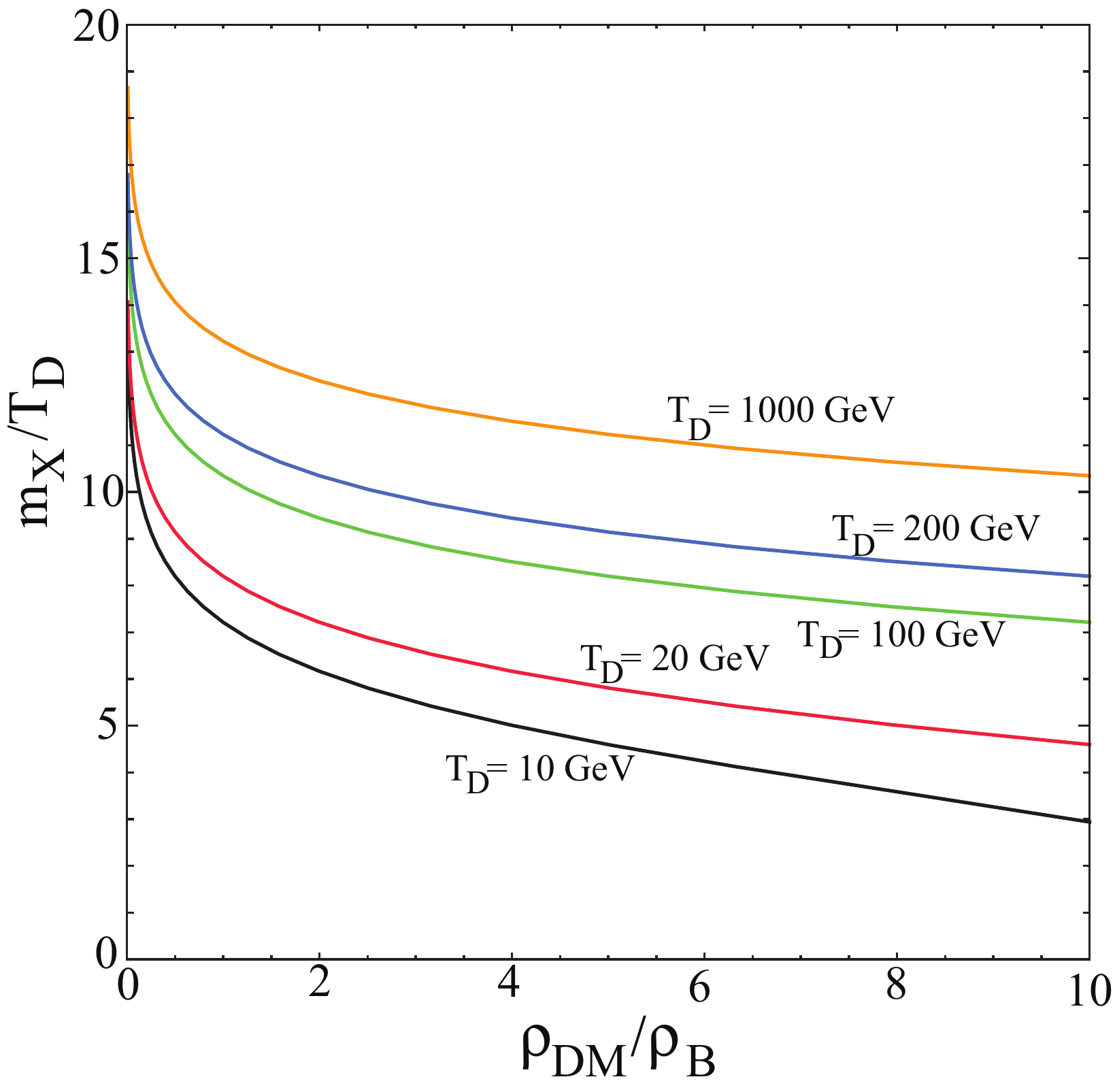}~~\includegraphics[width=0.4\textwidth]{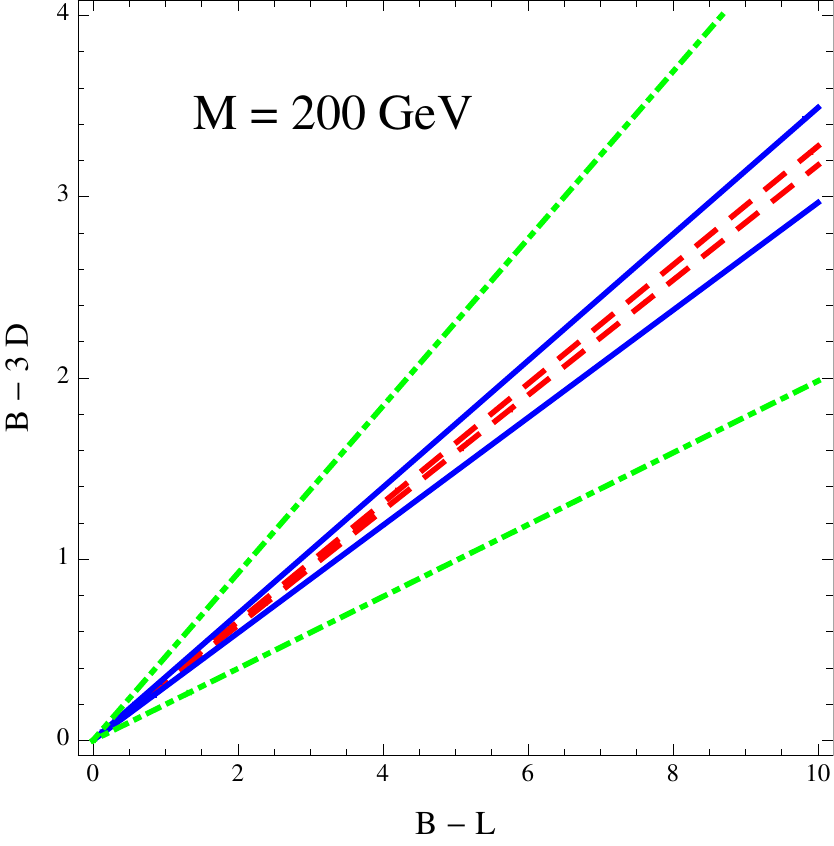}
\caption{{\em left}: $\rho_{DM}/\rho_B$ versus the DM mass to DM-number violation decoupling temperature ratio, $m_X/T_D$.  This plot is designed to illustrate how Boltzmann statistics with $m_X/T_D > 1$ can suppress the DM energy density, giving rise to the observed $\rho_{DM}/\rho_B$ even for heavy DM mass.  For decoupling of DM number violation at 200 GeV, the DM must be at least 2 TeV.  From \cite{Buckley:2010ui}.  {\em right}: In the absence of a large DM density suppression from a large DM to decoupling temperature ratio, cancellations in the sphaleron conserved quantities $B-L$ and $B-3 D$ can be used to achieve the correct density.  The dashed, solid and dot-dashed (red, blue, green) lines correspond to $\Omega_{DM}/\Omega_B = (1,5,25)$, with $m_X/T_D = 0.25$ and DM mass $m_X = 200$ GeV.  From \cite{Kribs:2009fy}.}
\label{suppression}
\end{center}
\end{figure}

On the other hand, one can use Boltzmann statistics and $D$ violation through the sphalerons to allow for a DM mass above $m_Z/2$, evading the $Z$-pole constraints while still being consistent with $\rho_{DM}/\rho_B \sim 5$.  In particular, if the DM becomes non-relativistic while sphalerons are still in thermal equilibrium, its number density is suppressed with respect to the natural scale by a factor \cite{Barr:1990ca,Barr:1991qn}
\begin{equation}
f(x) = \frac{3}{2 \pi^2}\int_0^\infty \frac{y^2 dy}{\cosh^2 \frac{1}{2} \sqrt{y^2 + x^2}},
\label{boltzmannsuppression}
\end{equation}
with $ x = M_*/T_D$ and $M_*$ the DM mass at the temperature $T_D$ where the electroweak sphalerons become inactive.  In this case, the ratio of DM to baryon densities is $\Omega_{DM}/\Omega_{b} = f(x) m_X D /m_p B$.  For the case where $B$ and $D$ are equal, $m_X/T_D$ is shown in Fig.~\ref{suppression} with $M_* = m_X$.  Since $T_D$ is typically on the order of 200 GeV, this implies that the DM mass must be about 2 TeV, and hence its mass cannot be generated by the Higgs vacuum expectation value alone.  Either the DM mass is generated primarily through some additional mechanism (such as the confinement of a technicolor gauge group), or one makes use of a cancellation between $B$,~$L$ and $D$ that lowers the DM number density.

To determine how a cancellation can occur in some cases, a standard calculation (which is reviewed in the appendix) \cite{Harvey:1990qw} is carried out on chemical potentials. Interactions give relations between the chemical potentials and $U(1)_{em}$ enforces electric neutrality.  For example, for the case that the DM is a bound state of two $SU(2)$ and electrically charged $\pm 1/2$ constituents, each of which have $D = 1/2$, one finds \cite{Kribs:2009fy}   
\begin{eqnarray}
B & = & \frac{ (36 f(x) + 4 f(x)^2) I_1 + (17 + 2 f(x))I_2}{17 + 113 f(x) + 13 f(x)^2} \\
L & = & \frac{ -(17+77 f(x) + 9 f(x)^2) I_1 + (17 + 2 f(x))I_2}{17 + 113 f(x) + 13 f(x)^2} \\
D & = &f(x) \frac{ (36 f(x) + 4 f(x)) I_1 - (111 +132 f(x))I_2}{51 + 339 f(x) + 39 f(x)^2}. 
\label{SphaleronRelations}
\end{eqnarray}
The resulting ratio of energy densities (when $f(x) \approx 1$) is
\begin{equation}
\frac{\Omega_{DM}}{\Omega_B} \simeq \frac{36 I_1 - 111 I_2}{51 I_2} \frac{m_X}{m_p}.
\end{equation}
The cancellation between $I_1$ and $I_2$ to obtain the observed density, when $m_X/m_p$ is quite large, is shown in Fig.~(\ref{suppression}), where it is seen that some tuning between $I_1$ and $I_2$ is necessary in order to obtain $\Omega_{DM}/\Omega_B \approx 5$.

Electroweak precision tests generally place strong constraints on models where the asymmetry is generated via sphalerons.  The reason is that $S$ and $T$ parameters strongly constrain additional electroweak states with chiral masses (see {\em e.g.} \cite{Kribs:2007nz}), and in order to generate an asymmetry in the dark sector via electroweak sphalerons, the DM must have chiral masses.  For example, the theory with a pure vector mass for the DM, $m_X \bar X X$, with $X$ and $\bar X$ fundamentals of $SU(2)$ but carrying opposite $D$ number, has no electroweak anomaly and hence does not couple to sphalerons.   

On the other hand one need not use electroweak sphalerons to transfer an asymmetry between sectors.  A new gauge group ${\cal G}$ can be introduced under which both DM number $D$ and $B$ and/or $L$ are anomalous.  This also allows one to easily evade precision electroweak and collider constraints on matter charged under $SU(2)_L$, while still providing for a mechanism to share the asymmetry between visible and dark sectors.  This mechanism was employed, for example, in \cite{Blennow:2010qp}, where the initial asymmetry was generated in $L$ via leptogenesis and then transferred via a new gauged horizontal symmetry $SU(2)_H$, under which $B$, $L$ and $D$ are anomalous, while $B-L-D$ is conserved.  A new anomalous dark gauge group ${\cal G}$ was  also employed in \cite{Dutta:2006pt,Dutta:2010va}, though in that case the asymmetry was generated from the out-of-equilibrium dynamics associated with a phase transition breaking ${\cal G}$, by analogy with electroweak baryogenesis.  The asymmetry is then transferred to the visible and DM sectors via decay of heavy states that obtained asymmetries via the phase transition for ${\cal G}$.

In the next section we review the second mechanism for sharing the asymmetry between sectors that is automatically free from constraints on electroweak sphaleron mediated models.

\subsection{Higher Dimension Operators and Renormalizable Interactions}
\label{subsec:highdimops}

All of the issues with electroweak sphaleron mediated ADM transfer mechanisms can be evaded if the transfer mechanism involves no SM quantum numbers.  The basic idea is to write down an interaction between $B-L$ in the SM and the global DM number, $D$.  Rapid scatterings through the interaction, inducing chemical equilibrium between the DM and SM, will cause an asymmetry in $B-L$ to be shared with the DM.  The precise nature of the interaction will determine how large the DM asymmetry is in comparison to the $B-L$ asymmetry.  At low temperatures the asymmetry sharing interactions must decouple, otherwise the asymmetry in the DM will be washed-out through the same interactions that created it.   

\begin{figure}
\begin{center}
\includegraphics[width=0.9\textwidth]{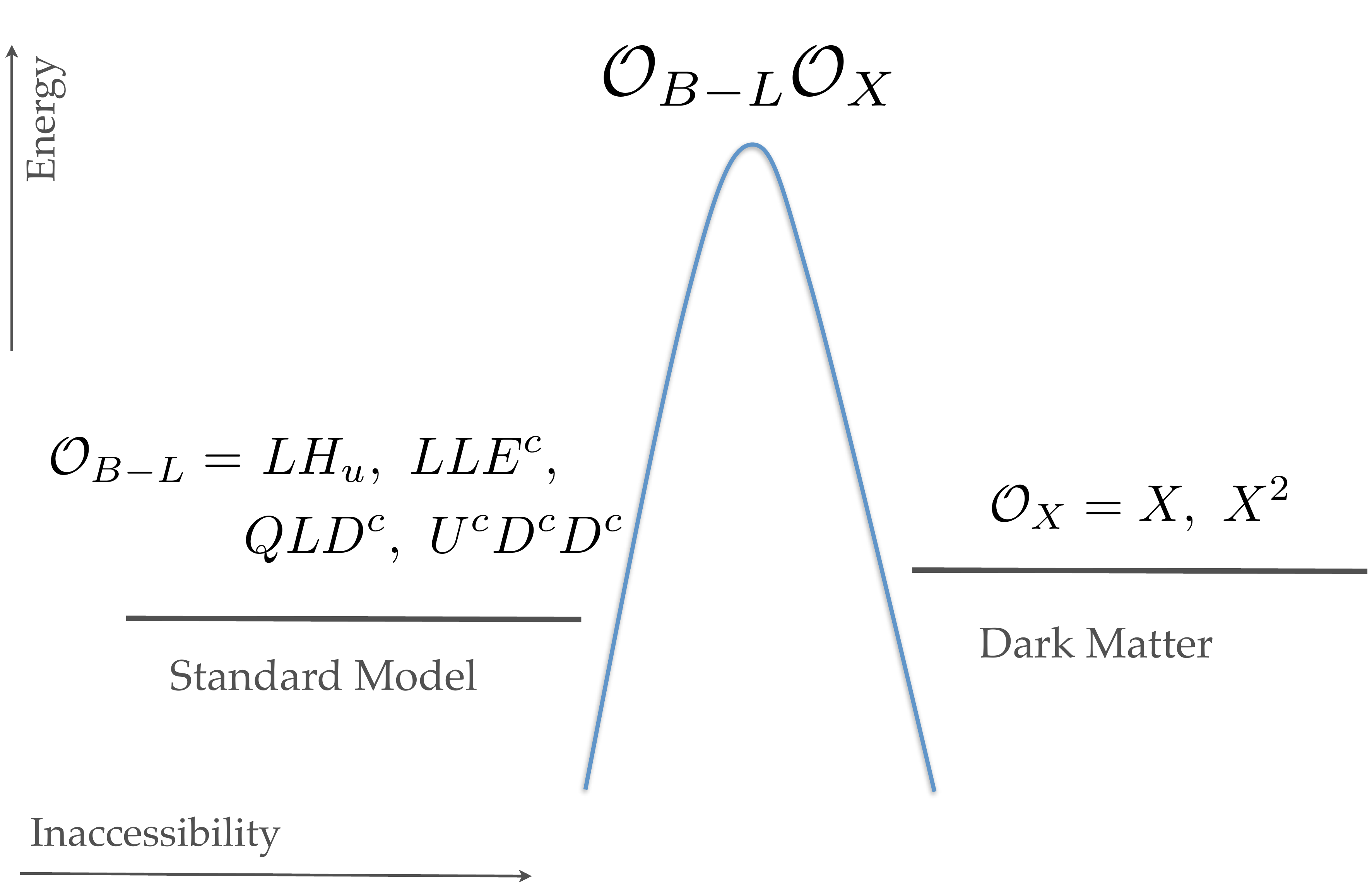}
\caption{A schematic of higher dimension ADM models \cite{Kaplan:2009ag}.  The $x$-axis represents the inaccessibility of the sector (the visible or dark sector) and the $y$-axis its energy, with the set-up inspired by Hidden Valley models \cite{Strassler:2006im}.  The mediator of the higher dimension operator, ${\cal O}_D {\cal O}_{B-L}$, presents a barrier between the two sectors.  At high temperatures the barrier is effectively removed though high energy interactions, allowing the $B-L$ and $D$ number to be shared between the two sectors.  At low temperatures, the barrier effectively freezes in the asymmetry between the two sectors.}
\label{ADMSchematic}
\end{center}
\end{figure}

A natural way for both the transfer and decoupling to occur is via higher dimension operators \cite{Kaplan:2009ag}, schematically shown in Fig.~\ref{ADMSchematic}.   A heavy field mediates the interaction, which is shown as a barrier between the visible and dark sectors in Fig.~\ref{ADMSchematic}.  At high temperatures, this barrier is not visible to the interactions, and the asymmetry is freely shared between the two sectors.  As the temperature in the early universe drops, the barrier becomes visible, and the interactions through the heavy mediator decouple, separately freezing in an asymmetry in each of the two sectors. 

On the SM side, the basic ingredient of the interactions as utilized in \cite{Kaplan:2009ag} is combinations of SM fields that carry no gauged SM quantum numbers, but carry $B-L$:
\begin{equation}
{\cal O}_{B-L} = u^c d^c d^c,~q \ell d^c,~\ell \ell e^c,~\ell H_u,
\label{SMops}
\end{equation}
where $u^c,~d^c,~e^c$ are right-handed anti-quarks and charged leptons, $q,~\ell$ are left-handed quark and lepton doublets, and $H_u$ is the up-type Higgs.
These operators can communicate their baryon or lepton chemical potential to DM fields via interactions with dark sector fields that carry DM number $D$.  The simplest combination of DM fields, $X$, is simply an operator carrying $n$ units of $D$ number:
\begin{equation}
{\cal O}_D = X^n.
\end{equation}
 In these models, since it carries no SM quantum numbers, the DM is quite unconstrained by the usual searches for new states (notably at colliders) and so can be quite light.  Thus no Boltzmann suppression via Eq.~(\ref{boltzmannsuppression}) is needed.

By positing an interaction between ${\cal O}_{B-L}$ and ${\cal O}_D$, the non-zero chemical potential in one sector is shared with the other sector.
Thus for example, we can write an interaction between the operators \cite{Kaplan:2009ag}
\begin{equation}
W = \frac{{\cal O}_D {\cal O}_{B-L}}{M^{m+n-3}},
\label{HigherDimOp}
\end{equation}
where $m$ is the dimension of ${\cal O}_{B-L}$.  Here we have written the interaction through a superpotential $W$ since these models were originally realized in the context of supersymmetry, though the theory need not be supersymmetric.   The lowest dimension operators in the supersymmetric theory take the form
\beq
W = \frac{X u^c d^c d^c}{M},~\frac{X q \ell d^c}{M},~\frac{\ell \ell e^c}{M},~y X L H.
\label{highdimops}
\eeq
The higher dimension operators can be implemented in the context of Grand Unified Theories \cite{Barr:2011cz} or in string theory \cite{Heckman:2011sw}. An operator similar to $L H X^2$ was also implemented in \cite{Foot:2003jt,Foot:2004pq}. 

One can re-phrase this discussion about interactions in terms of symmetries.  We take a convention where ${\cal O}_D$ has charge $+1$ under $D$.  When the interactions are rapid in the early universe, ${\cal O}_D {\cal O}_{B-L}$ explicitly breaks $B-L$ and $D$ down to an exact, diagonal $B-L+D$.  When the interactions decouple, $B-L$ and $D$ symmetries are each separately restored, freezing in the $B-L$ and $D$ numbers separately.

To determine the exact relation between the DM and baryon asymmetries, one must do a calculation that depends on the details of how the interaction decouples, as well as the spectrum of particles.  In addition to the $B-L$ and $D$ violation through the interaction of Eq.~\ref{HigherDimOp}, electroweak sphalerons violate $B+L$, which must be included as part of the calculation.  The relative abundance of $D$, $B$ and $L$ depends on linear chemical equilibrium equations, which can be easily solved using the methods outlined in \cite{Harvey:1990qw}, and which we review in appendix~\ref{subsec:distrib}.   The results depend on whether  the operator Eq.~(\ref{HigherDimOp}) decouples above or below the electroweak phase transition, since the equations being solved change (see the appendix for details).  If ${\cal O}_{B-L}$ carries $B-L$ number $p$, and the operator decouples above the electroweak phase transition when all minimal supersymmetric SM (MSSM) and $X$ particles are in thermal equilibrium, then \footnote{All the operators in Eq.~\ref{SMops} have $|p| = 1$}
\begin{equation}
m_X = \frac{79}{22} \frac{n}{|p|}\frac{\Omega_{DM}}{\Omega_B}\frac{B}{B-L}.
\end{equation}
On the other hand, if the operator decouples below the electroweak phase transition, when the top and all Higgs particles have already thermally decoupled, then
\begin{equation}
m_X = \frac{9}{2} \frac{n}{|p|}\frac{\Omega_{DM}}{\Omega_B}\frac{B}{B-L}.
\end{equation}
Depending on the details of the electroweak phase transition $B/(B-L) \sim 0.3$, with an uncertainty of approximately $15\%$.

The number of interactions available to communicate the asymmetry between sectors can be greatly increased if one adds additional fields.  For example, the Higgs can be employed \cite{Servant:2013uwa} with an extended $X$ sector, and the {\em renormalizable} interaction $X X_2 H$, where $X_2$ is an $SU(2)$ doublet, to transfer an asymmetry between the Higgs chemical potential (which may be non-zero before electroweak symmetry breaking) and the DM before electroweak symmetry breaking.  A related scenario instead makes use of an extended $X$ sector with lepton violating interactions to transfer the asymmetry \cite{Cui:2011qe}.  One can also employ renormalizable interactions through heavy vector-like fermions \cite{Dulaney:2010dj,Perez:2013nra}, and operators of the form $X u_R u'_L$, where $u'_L$ is a new field carrying $SU(3)$ quantum numbers.  In these cases, the interaction decouples as the heavy states become integrated out, at which point these models become like their higher dimension operator counterparts.

The renormalizable interaction $L H_u X$ has been effectively utilized in the context of asymmetric neutrino or sneutrino DM  \cite{Dodelson:1993je,Hooper:2004dc,Falkowski:2009yz,Haba:2011uz,Kang:2011wb,Okada:2012rm,Choi:2013fva}\footnote{In some of these models a Majorana mass for the neutrino was employed.  In this case, the DM asymmetry will be erased, in some cases causing a wash-out of the DM density itself.  We discuss further wash-out due to $D$ violation in Sec.~\ref{sec:oscillations}.  As a result, some of these models do not successfully achieve the DM density.}.  The $LH$ operator also plays an important role in leptogenesis based decay models, as we discuss next.

\section{Mechanisms for Generating a Primordial Dark and Baryon Asymmetry}
\label{sec:baryo}

In the previous section we focused on transfer of a pre-existing lepton or baryon asymmetry to the DM sector.  Since ADM models may simply transfer a pre-existing asymmetry, it can naturally be embedded with the standard mechanisms for generating the cosmological lepton or baryon asymmetry.  Here we discuss how these mechanisms can be extended to simultaneous generation of asymmetries in the DM and visible sectors  through a single source (cogenesis), or to generation of the asymmetry by the DM sector (darkogenesis) that is transferred to the SM using one of the two mechanisms described in the previous section.   For the purposes of this review, darkogenesis and cogenesis are distinguished by whether the model has distinct mechanisms for transferring and generating the asymmetry.

The usual Sakharov criteria for baryogenesis is that the mechanism must
\begin{enumerate}
\item furnish a departure from thermal equilibrium;
\item have sufficient $C$ and $CP$ violation;
\item violate baryon number.
\end{enumerate}
The SM rather famously fails to meet the Sakharov criteria sufficiently to create the observed baryon asymmetry, both because $CP$ is not amply violated in the SM, and because the electroweak phase transition does not feature a suitably strong departure from thermal equilibrium.  In both darkogenesis and cogenesis the first two criteria are met as in baryogenesis or leptogenesis, while the last criterion may be evaded if the DM is itself viewed as carrying baryon number.

\subsection{Cogenesis}
\label{subsec:cogenesis}

Cogenesis models currently in the literature fall into three distinct categories.   {\em (i):} Decay models, where a parent particle simultaneously decays to the visible and hidden sectors.  {\em (ii):} Affleck-Dine mechanisms, where a flat direction carrying both $B-L$ and $D$ simultaneously generates both asymmetries.  {\em (iii):} Electroweak baryogenesis models that are extended to include the DM sector.  We review each of these classes.


\subsubsection{Decay}
\label{subsubsec:decay}

In standard baryo- or lepto-genesis decay models (see, {\it e.g.} \cite{Riotto:1999yt} for a review), an out-of-equilibrium, $CP$-violating and $L$ or $B$ violating decay generates a matter-anti-matter asymmetry.  The first of the Sakharov criteria is satisfied by the lifetime of the particle, $\tau$, exceeding the age of the universe (set by the Hubble expansion) when the universe's temperature drops below the mass of the decaying particle.  $C$ and $CP$ violation is usually achieved through the interference of tree and one-loop diagrams in the decay, while $B$ and $L$ can be explicitly violated in the Lagrangian through the decaying particle's interactions.

The decay of the parent can easily generate an asymmetry in the DM sector at the same time as a lepton or baryon asymmetry.  This is most easily illustrated through an extension of the classic leptogenesis mechanism \cite{Fukugita:1986hr}.  In the standard leptogenesis model, a sterile neutrino $N$ with an $L$ violating Majorana mass has out-of-equilibrium and $CP$ violating decays to leptons.  

\begin{figure}
\begin{center}
\includegraphics[width=0.5\textwidth]{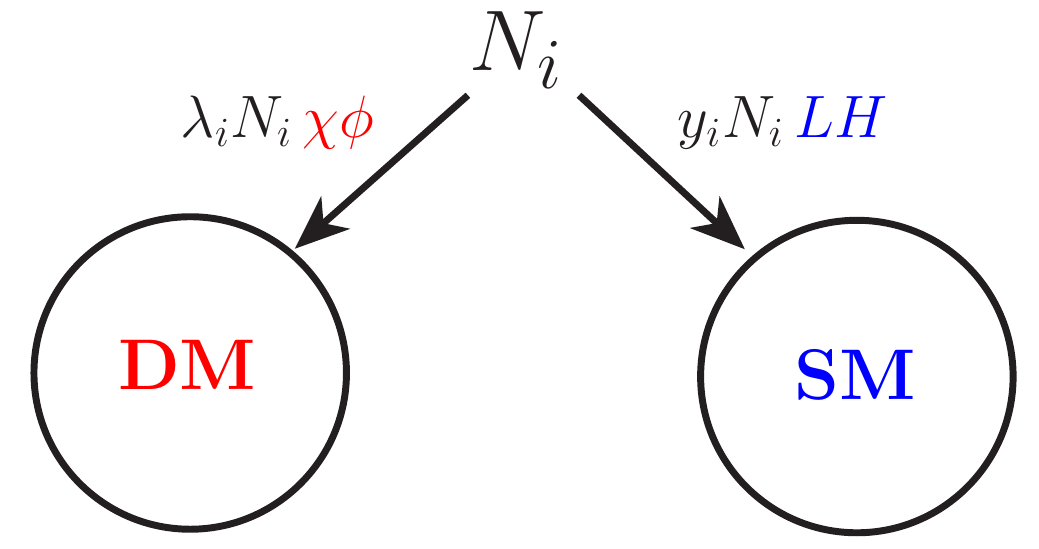}
\caption{A schematic of leptogenesis ADM models.  Out-of-equilibrium and CP violating decays of a sterile neutrino into both the SM and DM sectors give rise to an asymmetry in both sectors.  From \cite{Falkowski:2011xh}.}
\label{Leptogenesis}
\end{center}
\end{figure}

$N$ need not generate a lepton asymmetry alone, however.  $N$ may also generate a DM asymmetry if $N$ also has $CP$-violating decays to DM fields $\chi,~\phi$ at the same time. The extension we describe is shown schematically in Fig.~\ref{Leptogenesis} \cite{Falkowski:2011xh}.  The Lagrangian for the theory is
\beq
{\cal L} = M_i N_i^2 + y_i N_i LH + \lambda_i N_i \chi \phi,
\eeq
where multiple sterile neutrino species $N_i$ are necessary in order to generate the CP violation in the decay, as in the standard mechanism.   Also note an important connection to the higher dimension transfer mechanism: if we integrate out the sterile neutrino $N$, we generate an operator of the form $\chi \phi LH/M$, that is very similar to a $X^2 LH/M$ higher dimension interaction introduced in Eq.~\ref{HigherDimOp}.

One of the unique features of leptogenesis based ADM models is that, since $N$ is itself $B-L$ and $D$ violating, the asymmetries in the $D$ and $B-L$ sectors need not be the same.  In fact, the asymmetries that are generated in each sector are proportional to the $CP$ violation in the decays of $N_i$ to each sector \cite{Falkowski:2011xh}:
\beq
\epsilon_\chi = \frac{\Gamma(N_1 \rightarrow \chi \phi) - \Gamma(N_1 \rightarrow \bar{\chi}\phi^\dagger)}{\Gamma_{N_1}},~~~~~~\epsilon_l = \frac{\Gamma(N_1 \rightarrow l h) - \Gamma(N_1 \rightarrow \bar{l} h^\dagger)}{\Gamma_{N_1}}.
\eeq
The ratios of the asymmetries are parametrically set by the ratios of the Yukawa couplings along with the phases $\phi_l,~\phi_\chi$ in $y_i$ and $\lambda_i$:
\beq
\frac{\epsilon_l}{\epsilon_\chi} \simeq \frac{2 r \sin(2 \phi_l) + \sin(\phi_l + \phi_\chi)}{2 r^{-1} \sin(2 \phi_\chi) + \sin(\phi_l + \phi_\chi)},
\eeq
$r = y_1 |y_2|/\lambda_1 |\lambda_2|$.    Since {\em a priori} the Yukawas and phases in the SM and DM sectors need not be the same, {\em the DM mass in this model need not be close to the proton mass.}
This asymmetry may be partially washed out by $2 \leftrightarrow 2$ processes that re-equilibrate part of the asymmetries in the DM and visible sectors, such as $\phi \chi \leftrightarrow \bar{L} H^\dagger$ and $\phi \chi \leftrightarrow L H$.  As in conventional models of leptogenesis, these lepton and DM violating processes must be out-of-equilibrium to avoid washout of the asymmetries.  This basic leptogenesis scenario can be realized in many different models, and we refer the reader to \cite{Cosme:2005sb,An:2009vq,Gu:2009hj,Gu:2009yy,Chun:2010hz,Gu:2010yf,Davoudiasl:2011aa,Chun:2011cc,Arina:2011cu,Arina:2012aj,Choi:2012ba,Arina:2012fb,Kuismanen:2012iz,Canetti:2012kh,Canetti:2012vf} for further details.

This leptogenesis model makes use of, when $N$ is integrated out, the $L H {\cal O}_D$ ADM type of operator of Eq.~\ref{HigherDimOp}.  Other interactions can be used in out-of-equilibrium decay scenarios \cite{Davoudiasl:2010am,Gu:2010ft,Feng:2013wn}.  For example, the ${\cal O}_D u^c d^c d^c$ operator can be implemented in a baryogenesis model with the Lagrangian \cite{Davoudiasl:2010am}
\begin{equation}
{\cal L} = \frac{\lambda_i}{M^2} S_i u^c d^c d^c  +  \zeta_i \bar{S}_i \chi \phi + M_S^i \bar{S}_i S_i,
\label{hylogenesis}
\end{equation} 
where multiple $S_i$ are necessary to generate the CP-violation in the decay, similar to the leptogenesis model.  Here, $S$'s out-of-equilibrium decays generate an asymmetry in baryon number, as well as the hidden sector fields $\chi$ and $\phi$.  Since $S$ has a Dirac and not Majorana mass, however, the asymmetry generated in $B-L$ and in $D$ is {\em equal and opposite}, which is different from the leptogenesis scenario.
The most interesting phenomenological consequence of this theory is that it allows for DM catalyzed proton decay \cite{Davoudiasl:2011fj}.  For example, $\bar{\chi} p \rightarrow K^+ \phi$ or $\phi^* p \rightarrow \chi K^+$.    In a related model, instead of a small $CP$ violating parameter $\epsilon$ giving rise to the small baryon asymmetry $\eta \sim 10^{-10}$, moduli can effectively dilute an ${\cal O}(1)$ asymmetry \cite{Allahverdi:2010rh}.   Lastly, the ``asymmetric freeze-in'' scenario \cite{Hall:2010jx} is closely related to the previous decay scenarios.  For example, one can use the $W = \lambda_i L_i H X$ operator alone, with CP-violating decays of the charginos and neutralinos to a SM lepton and the DM.  It has been argued \cite{Hook:2011tk}, however, that unitarity and CPT lead to cancellations that prevent the transfer of baryon number, so that this model is not ultimately successful.  

When the DM sector is thermalized, the dominant symmetric component must annihilate, leaving only the asymmetric abundance.  We return in Sec.~\ref{sec:hidden sectors} to discussing this process of freezing-out the asymmetric abundance.  Another class of models \cite{Kitano:2004sv,Kitano:2005ge} exchanges the annihilation of the thermal symmetric abundance for a two state decay process to the DM.  In these models the parent particle $P$ decays to a quark and a messenger particle $M$.  The key difference between this model and the decay models we have already discussed is that this intermediate mass messenger particle (which carries an asymmetry itself) first annihilates away its thermal symmetric abundance, leaving only the (small) asymmetric abundance, {\em before} it decays down to the DM particle $X$.  Thus $M$ must be relatively long lived, and its decays generally occur through higher dimension operators.  Related models \cite{Thomas:1995ze,Unwin:2012rp} start with an asymmetry in $P$, which annihilates away its thermal symmetric abundance.  $P$ then decays to baryons and leptons in SM sector (which carry the parent asymmetry), and to the DM sector.  The DM itself need not be asymmetric, though its abundance was set by the asymmetry in $P$.  A similar case where the DM is itself not asymmetric, but its density was set by the asymmetry in $P$ was considered for gravitino DM in \cite{Kang:2011ny}.

\subsubsection{Affleck-Dine Mechanisms}

Supersymmetric ADM models via the higher dimension operators of Eq.~(\ref{HigherDimOp}) make a natural playing field for Affleck-Dine (AD) baryogenesis \cite{Affleck:1984fy} that simultaneously generates the DM and baryon asymmetries.  The AD baryogenesis mechanism  makes use of supersymmetric flat directions carrying $B-L$ to satisfy the last Sakharov criterion of baryon violation. Out-of-equilibrium dynamics is provided by the end of inflation and the re-heating of the universe, while $CP$-violating phases in the scalar fields satisfy the second Sakharov criterion.  

We briefly review how the AD mechanism works before describing its applications to ADM.  In classic AD, inflation induces supersymmetry breaking terms proportional to the Hubble parameter $H$ that drive a $B-L$ carrying field $\phi$ to take a non-zero vev.  These terms are of the type \cite{Dine:1995uk,Dine:1995kz}\footnote{The stabilization away from the origin in field space may also be generated by $A$-terms, as we will see in an explicit example below.}
\begin{equation}
V_{soft} = \sum_\phi (a_\phi m^2 + b_\phi H^2)|\phi|^2.
\end{equation} 
If $b_\phi < 0$, the Hubble induced terms force $\langle B - L \rangle$ to take a non-zero value.  As $H$ drops below $m$ during re-heating, the field regains a global minimum at the origin ($\langle B - L \rangle = 0$) in field space.  When this happens, a torque is applied to $\phi$ -- this happens through the $A$-terms since these terms carry phases, which can thus change $\phi$.  As the field spirals in towards the new global minimum at the origin in field space, a non-zero $n_{B-L}$ results, since
\begin{equation}
n_\phi = j^0 = i(\phi \dot{\phi}^\dagger - \phi^\dagger \dot{\phi}) = r_\phi^2 \dot{\theta}_\phi,
\end{equation}
where $r_\phi$ is defined by $\phi = \frac{1}{\sqrt{2}} r_\phi e^{i \theta_\phi}$.
The net asymmetry that is generated during re-heating is
\begin{equation}
\eta_b = \frac{n_b}{s} \sim \frac{n_b}{\rho_\chi/T_R},
\end{equation} 
where $T_R$ is the re-heat temperature and $\rho_\chi$ is the inflaton density when Affleck-Dine baryogenesis occurs.

AD cogenesis extends the generation of $B-L$ to a simultaneous generation of $B-L$ and $D$ making use of supersymmetric flat directions that carry both global quantum numbers.  In the context of higher dimension ADM models, the key aspect of AD cogenesis \cite{Cheung:2011if} is the observation that the higher dimension operator of Eq.~\ref{HigherDimOp},
$
{\cal O}_{B-L} {\cal O}_D,
$
when ${\cal O}_{B-L}$ carries charge -1 under $U(1)_{B-L}$ and ${\cal O}_D$ charge +1 under $U(1)_D$, preserves $B-L+D$.  Because $B-L+D$ is preserved along the direction in field space employed for AD cogenesis, 
\begin{equation}
-n_{B-L} = n_D \neq 0.
\end{equation} 
 The $B-L$ and $D$ carrying fields are stabilized at non-zero vevs carrying $B-L$ and $D$ (but preserving $B-L+D$) via the soft term
\beq
V_{\rm soft} = (f m+g H)\frac{{\cal O}_{B-L} {\cal O}_D}{M^{d-4}},
 \eeq
 where $d$ is the dimension of $
{\cal O}_{B-L} {\cal O}_D.
$
Using an impulse approximation when $H = |f| m / |g|$ to calculate the asymmetry generated as the pseudo particle is forced out of its false minimum, we find \cite{Cheung:2011if}
\beq
-n_{B-L} = n_X \sim \frac{\mbox{arg}(f/g)g |{\cal O}_{B-L}||{\cal O_D}|}{M^{d-4}}.
\eeq
It can be shown (see \cite{Cheung:2011if} for details) that for typical parameters of the models an asymmetry consistent with the observed asymmetry may be obtained.

Other mechanisms exist within AD models to relate the DM and baryon numbers \cite{Enqvist:1997si,Enqvist:1998en,Suematsu:2005zc,Roszkowski:2006kw,Shoemaker:2009kg,Bell:2011tn,vonHarling:2012yn,Graesser:2011vj}.  One of these is to make use of the baryon or lepton number stored in $Q$-balls \cite{Kusenko:1997ad,Kusenko:1997hj,Kusenko:1997zq,Kusenko:1997si}.  $Q$-balls are solitonic configurations that arise along supersymmetric flat directions when the potential is flatter than quadratic, as is usually the case when supersymmetry is broken via gauge mediation \cite{Enqvist:1997si} (as well as in some gravity mediated scenarios).  $Q$-ball solutions arise when $U(\phi)/|\phi|^2$, where $U(\phi)$ is the potential for a globally symmetric complex scalar field $\phi$, has a global minimum at non-zero $\phi$. Because the potential is flatter than quadratic, fluctuations in the scalar field experience a negative pressure and grow exponentially, until the fluctuation becomes non-linear and collapses to form a $Q$-ball.

The baryon- or lepton-number carrying Q-ball decays to ordinary neutralino dark matter \cite{Enqvist:1998en}, as well as baryons or leptons.  The $Q$-ball condensate must decay after electroweak sphalerons have shut-off, otherwise the combination of $B+L$ violation by the sphalerons and the $B$ or $L$ violation along the $B$ or $L$ carrying flat directions washes out the asymmetry.  The baryon to DM ratio in this model is 
\begin{equation}
r_B = \frac{\rho_B}{\rho_{DM}} = \frac{m_n}{N_\chi f_B m_\chi},
\end{equation}
where $f_B$ is the fraction of the total baryon number that is trapped in the $B$-balls, and $N_\chi$ is the number of neutralinos produced per unit of $B$.  Related models promoted axino \cite{Roszkowski:2006kw} or gravitino \cite{Shoemaker:2009kg,Doddato:2011hx}, instead of neutralino, DM due to the LEP constraints on neutralinos (we note however that in general MSSM models the LEP constraints on light neutralino DM can be evaded).

Rather than making use of $Q$-balls to store some of the baryon asymmetry until after electroweak sphalerons have shut off the $B$ and $L$ violation in the SM, in some cases the condensate density may be of the same size as the baryon or lepton asymmetry generated by the Affleck-Dine mechanism \cite{McDonald:2006if}. 

The AD mechanism then transfers a fraction of the total energy density in the condensate, $\Phi$, into a baryon asymmetry according to the relation
\beq
n_B = \frac{f_A}{4} \frac{\rho_{\Phi 0}}{m_\Phi}.
\eeq  
$\rho_{\Phi 0} = m_\Phi^2 \phi(t_0)^2/2$ is the energy density in the oscillating field along the flat direction, and $f_A$ parameterizes the CP-violation (due to an $A$-term in the case of  \cite{McDonald:2006if}).

If this energy density (via an $(H_u L)^2$ flat direction in the case of \cite{McDonald:2006if}) is comparable to the DM density (via a sterile sneutrino condensate $N$), the baryon-DM coincidence is resolved:
\beq
\frac{\Omega_B}{\Omega_{DM}} = \frac{f_A}{4} \frac{m_n}{m_\Phi}\frac{\rho_{\Phi 0}}{\rho_{N 0}}.
\eeq
That $\rho_{\Phi 0}$ and $\rho_{N 0}$ are comparable is plausible if the superpotential terms lifting the sneutrino and $(H_u L)^2$ flat directions are similar,
\beq
W_{lift} = \lambda_N \frac{N^4}{M} + \lambda_\Phi \frac{\Phi^4}{M},
\eeq
such that $\lambda_\Phi \approx \lambda_N$.

Before we leave this section, we note that anti-baryon number may be stored in dense nuggets, analogous to $Q$-balls, but formed at the QCD phase transition \cite{Zhitnitsky:2002qa,Oaknin:2003uv}.  Because the density of these nuggets is approximately the same as ordinary hadronic matter, the baryon-DM coincidence is solved.  The phenomenology of this scenario is rich \cite{Zhitnitsky:2006vt,Lawson:2012zu,Lawson:2013bya}, though distinct from the particle ADM scenario that is the focus of this review.

\subsubsection{Electroweak Cogenesis}
\label{susubsec:electorweakcogeneis}

Lastly, the generation of the DM asymmetry can be directly tied to electroweak baryogenesis, as in electroweak cogenesis \cite{Cheung:2013dca}.  Electroweak baryogenesis makes use of a first order phase transition in the Higgs sector to provide the needed out-of-equilibrium dynamics, interactions of the Higgs boson to accommodate $C,~CP$ violation, and electroweak sphalerons to contribute $B$ violation, thus satisfying all the Sakharov criteria (see \cite{Cohen:1993nk,Morrissey:2012db} for a review of electroweak baryogenesis).  

If the global $D$ symmetry is spontaneously broken in the early universe through the vev of a scalar field, an asymmetry in $D$ can be generated at the same time as the baryon asymmetry.  The electroweak phase transition may then trigger a restoration of the $D$ symmetry, and if this transition is sufficiently first order, the DM asymmetry is frozen in. The concrete scenario considered in \cite{Cheung:2013dca} features two electroweak Higgs doublets ($H_1,~H_2$) to obtain a sufficiently first-order phase transition for the baryon asymmetry.  There are also two dark scalars, $S_1$ and $S_2$ carrying $D$, one of which ($S_2$) has a vev before the electroweak phase transition that breaks $D$.  As one of the electroweak Higgses obtain a vev, this changes the potential of $S_1$ and $S_2$, such that $S_2$ has its symmetry restored by obtaining zero vev.  As this happens, $C,~CP$ violating interactions on the bubble wall of $S_2$ generates a $D$ asymmetry.  While this simultaneously generates a DM and baryon asymmetry at the electroweak phase transition, the phases and dynamics of the electroweak and $S_2$ phase transitions need not be identical, implying different asymmetries in the two sectors.

\subsubsection{Cogenesis and the WIMP Miracle}

As discussed in the introduction, the WIMP miracle is the observation that freeze-out through weak scale annihilation cross-sections gives rise to the observed relic abundance.  ADM in most cases forgoes this miracle, instead driving the DM density via a baryogenesis related mechanism.  Conversely, the WIMP miracle may be preserved, and the baryon asymmetry set via the WIMP miracle, an idea considered in \cite{Cui:2011ab,McDonald:2011sv,Cui:2012jh,Bernal:2012gv,Bernal:2013bga}.  In these models, the DM is not necessarily asymmetric, though baryon violating interactions in annihilations set the DM density.

For example, in \cite{Cui:2012jh}, the WIMP miracle was preserved via a metastable particle, which freezes out having its relic density set by annihilation in the usual way.  Then the parent $P$ has $CP$ and $B$ violating decays that generate the baryon  asymmetry, so that $\Omega_B = \epsilon m_p/m_X \Omega_{DM}$, where $\epsilon$ is the CP violating phase in the decay and $m_p$ is the parent mass.  If the decays to $X$ are $CP$-violating, an asymmetry may also be generated in the DM sector.  The degree to which $m_p$ and $m_X$ are of the same size depends on the strength of the $CP$-violation in the decay to the DM.

On the other hand, the baryon asymmetry can be set directly in baryon violating annihilations of the DM, as was considered in \cite{Farrar:2005zd,Cui:2011ab,McDonald:2011sv}.   If the asymmetry in these annihilations is $\epsilon$, then the asymptotic abundance $Y \equiv n/s$ is $Y_B = \frac{\epsilon}{2}\left[ Y_X(x_{washout})-Y_X(\infty) \right]$, where $x_{washout} = m_X/T_{washout}$ is when washout processes freeze out and $Y_X(\infty)$ is the DM density today.  This equation is understood as follows: after washout decouples, subsequent WIMP annihilations (which are of size $Y_X(x_{washout})-Y_X(\infty)$) generate a baryon asymmetry of size $\epsilon$ \cite{Cui:2011ab}.

\subsection{Dark Baryogenesis}
\label{subsec:darkogenesis}

We have already discussed the case where ADM is embedded in standard lepto- or baryo-genesis and then transferred, as well as the case where DM and visible asymmetries are generated at the same time.  The last case is dark baryogenesis (or darkogenesis), where an asymmetry is generated in the DM sector and then transferred, either by sphalerons or through interactions.  Both decay and Affleck-Dine mechanisms (which we only discussed in the context of cogenesis) could be implemented in the DM sector alone and then transferred via sphalerons or interactions, such as in \cite{Haba:2010bm}.  There are two additional examples in the literature of an asymmetry being generated in the DM sector and transferred, models that mimic electroweak baryogenesis\footnote{We also reviewed electroweak based mechanisms in the cogenesis section~\ref{susubsec:electorweakcogeneis}.  The mechanism is sufficiently different when carried out in the DM sector alone that we review it separately here.} or spontaneous baryogenesis in the DM sector.   We review each case.

\subsubsection{Dark ``Electroweak Baryogenesis''}
 
In a similar fashion to electroweak baryogenesis, a dark Higgs may have  $C,~CP$ violating interactions and a first-order phase transition, and the DM sector may have a dark non-Abelian gauge group under which DM number $D$ has a chiral anomaly so that dark sphalerons violate $D$.  Because the DM sector dynamics (including the means by which the DM mass itself is set) is still unknown, significant freedom remains to generate the needed out-of-equilibrium dynamics, as well as the needed $C$ and $CP$ violation.  In fact, most of the constraints in models of darkogenesis are derived from the mechanism that subsequently transfers the asymmetry to the visible sector, rather than by the asymmetry generation itself.

\begin{figure}
\begin{center}
\includegraphics[width=0.8\textwidth]{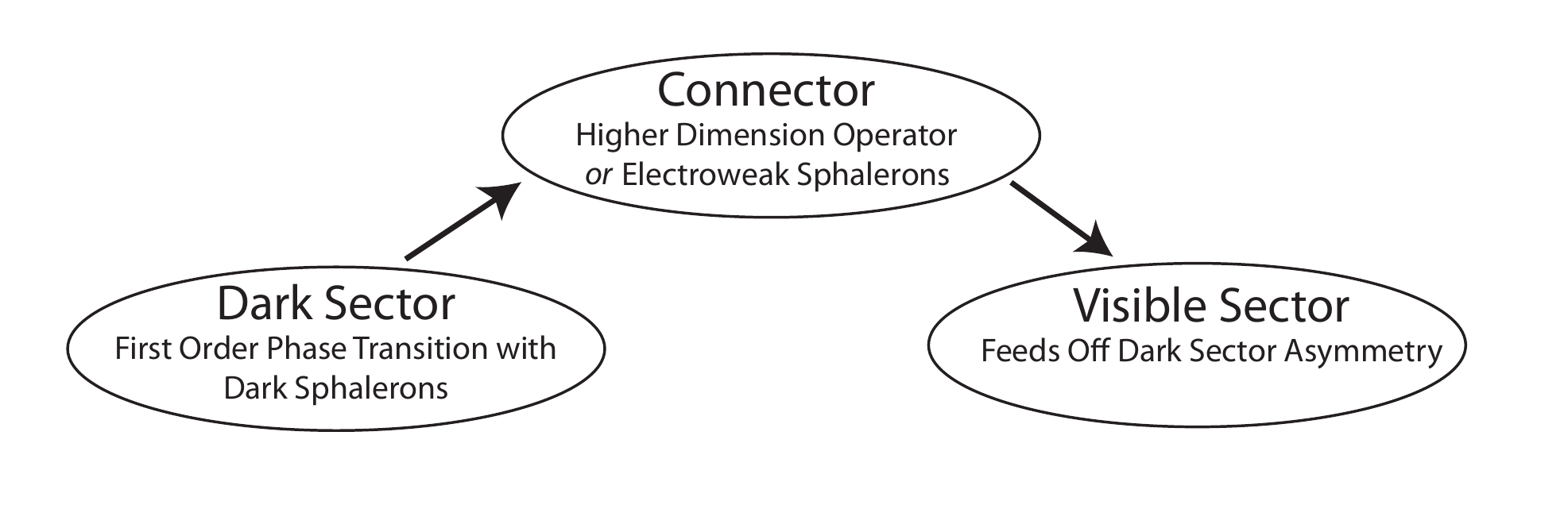}
\caption{Schematic of darkogenesis models.  Darkogenesis makes use of dynamics in the dark sector (with a dark Higgs and dark sphalerons) to generate a dark asymmetry that is then transferred to the baryons via a connector sector.  Figure from \cite{Shelton:2010ta}.}
\label{darkogenesis}
\end{center}
\end{figure}

Viable models of darkogenesis can be found in \cite{Shelton:2010ta,Dutta:2010va,Petraki:2011mv,Walker:2012ka}, and a schematic is shown in Fig.~\ref{darkogenesis}.   
To be explicit, darkogenesis works as follows.
\begin{itemize}
\item A dark non-Abelian gauge group $SU(N)_D$ is introduced.  If $D$ is anomalous under $SU(N)_D$ (exactly the way $B$ and $L$ are anomalous under SM $SU(2)$), then dark sphalerons can generate a non-zero $D$.
\item $SU(N)_D$ undergoes a first-order symmetry-breaking transition via a dark Higgs, which generates the needed out-of-equilibrium dynamics to give a dark asymmetry.  Fields carrying $D$ experience $C,~CP$ violation in their interactions with the dark Higgs.  In this way, the visible sector has its asymmetry generated via the dark sector. 
\item This asymmetry is transferred to the visible sector by a messenger sector, which can be via renormalizable or higher dimension interactions, or via electroweak sphalerons.  The transfer mechanisms discussed in Sec.~\ref{sec:models} thus directly apply once the asymmetry in $D$ has been generated.
\end{itemize}

The major question then is what the simplest DM sector is that could generate such dynamics.   The basic requirement is that the global $D$ symmetry be anomalous under $SU(N)_D$.  A simple Lagrangian that does this is
\begin{equation}
{\cal L} = -\frac{m^2 |H_D|^2}{2} + \frac{\lambda}{4} |H_D|^4 +y_{i} L_D H_D \bar{X}_i,
\end{equation}
where $i=1,~2$.  $L_D$ and $X$ carry $D$ number $+1$, but only $L_D$ is charged under $SU(N)_D$ making $D$ number anomalous under $SU(N)_D$.  Thus $SU(N)_D$ sphalerons violate $D$ number and can generate an asymmetry.  $H_D$ is a fundamental scalar boson under the $SU(N)_D$ that breaks $SU(N)_D$, driving the dark phase transition to the broken phase by analogy with the SM Higgs.   Because the parameters of the potential of $H_D$ are completely free, it is straightforward to generate a first-order phase transition, with sufficient $CP$ violation.  

Since a new fundamental scalar has been introduced, it is appealing to embed this model within supersymmetry to generate the needed GeV mass scale naturally, and give rise to the DM mass itself.  We return in Sec.~\ref{sec:hidden sectors} to concrete supersymmetric models that do this.  In the meantime, we next describe the transfer mechanisms within darkogenesis.

The simplest way to transfer the $D$ asymmetry is via higher dimension operators, making use of the operators in Eq.~\ref{HigherDimOp}.  A less trivial case is transfer via electroweak sphalerons.  To make use of the electroweak sphalerons to transfer the asymmetry, not only $B$ and $L$ must be anomalous under $SU(2)$, but also $D$.  The basic reason it is non-trivial to simultaneously violate all three with electroweak sphalerons is that anomaly cancellation plus mass generation forces a non-trivial structure in the messenger sector.  The simplest Lagrangian in the messenger sector, which transfers the asymmetry, is
\begin{equation}
{\cal L}_M = y_{eM+} L_M^+ H {e_M^c}^+ + y_{eM-} L_M^- H^\dagger {e_M^c}^- + y_{XM+} L_M^+ H {\bar{X}^i_M} + y_{XM-} L_M^- H^\dagger {\bar{X}^i_M}. 
\end{equation}  
The charges of these fields are given in Table~\ref{table:mess}. 
With the introduction of the messenger fields, the electroweak sphalerons now violate $B+L+\frac{N_D}{N_g}D$, where $N_D$ is the number of messenger electroweak doublets, so that electroweak sphalerons can efficiently reprocess a $U(1)_D$ asymmetry into $B$ and $L$.

\begin{table}
\begin{center}
\begin{tabular}{cccc}
\hline \hline
& $SU (2)_L$ & $U(1)_Y $ & $U(1)_{D} $\\
\hline

$L_{M}^\pm$      & 2 & $\pm\frac{1}{2} $ & $1$\\
${e_{M}^c}^\pm$     & 1 & $\mp 1$               & $-1$ \\
$\bar{X}_{M}^i$     & 1 & $0$               & $-1$ \\
\hline \hline
\end{tabular}
\caption{A minimal dark messenger
sector\label{table:messengercontent}.  Anomaly cancellation is
achieved via mirror fermions with the same $U(1)_{D}$ charge, but
opposite hypercharge.  There are two sterile states,
$\bar{X}^{1,2}_M$.
\label{table:mess}
}
\end{center}
\end{table}

\subsubsection{Spontaneous Dark Baryogenesis}

Instead of a dark phase transition (by analogy with electroweak baryogenesis), spontaneous baryogenesis \cite{Cohen:1988kt,Cohen:1991iu,Cohen:1994ss} can be employed to generate a DM asymmetry that is then transferred to $B$ and $L$.
In spontaneous baryogenesis a spontaneous violation of $CPT$ and $T$ replaces the usual $C$ and $CP$ violation of the Sakharov Criterion.  In spontaneous baryogenesis, a term in the Lagrangian is posited:
\beq
{\cal L} = \frac{\partial_\mu \phi}{f} J^\mu_B ,
\eeq
where $J^\mu_B$ is the baryon current.  If the scalar field obtains a vev, $\partial_\mu \phi = (\dot{\phi},{\bf 0})$, $CPT$ and $T$ are spontaneously broken, generating a chemical potential $\mu_B = \dot{\phi}/f$ and a net baryon asymmetry
\beq
n_B = n_b - n_{\bar b} = \frac{g_B \mu_B T^2}{6},
\eeq
where $g_B$ is the number of baryonic degrees of freedom.

Spontaneous baryogenesis can be carried out instead in the DM sector and then transferred to the visible sector, typically via the higher dimension operators of Sec.~\ref{subsec:highdimops}.  This mechanism was called ``spontaneous cogenesis,'' \cite{MarchRussell:2011fi,Kamada:2012ht} though the asymmetry is driven by the DM sector and then transferred, rather than being driven by both sectors.  If the DM sector decouples from the transfer operator before spontaneous baryogenesis has completed in the hidden sector, different asymmetries in the dark and visible sectors may result.

\section{Annihilating the Thermal Relic Abundance}
\label{sec:annihilation}

\begin{figure}
\begin{center}
\includegraphics[width=0.5\textwidth]{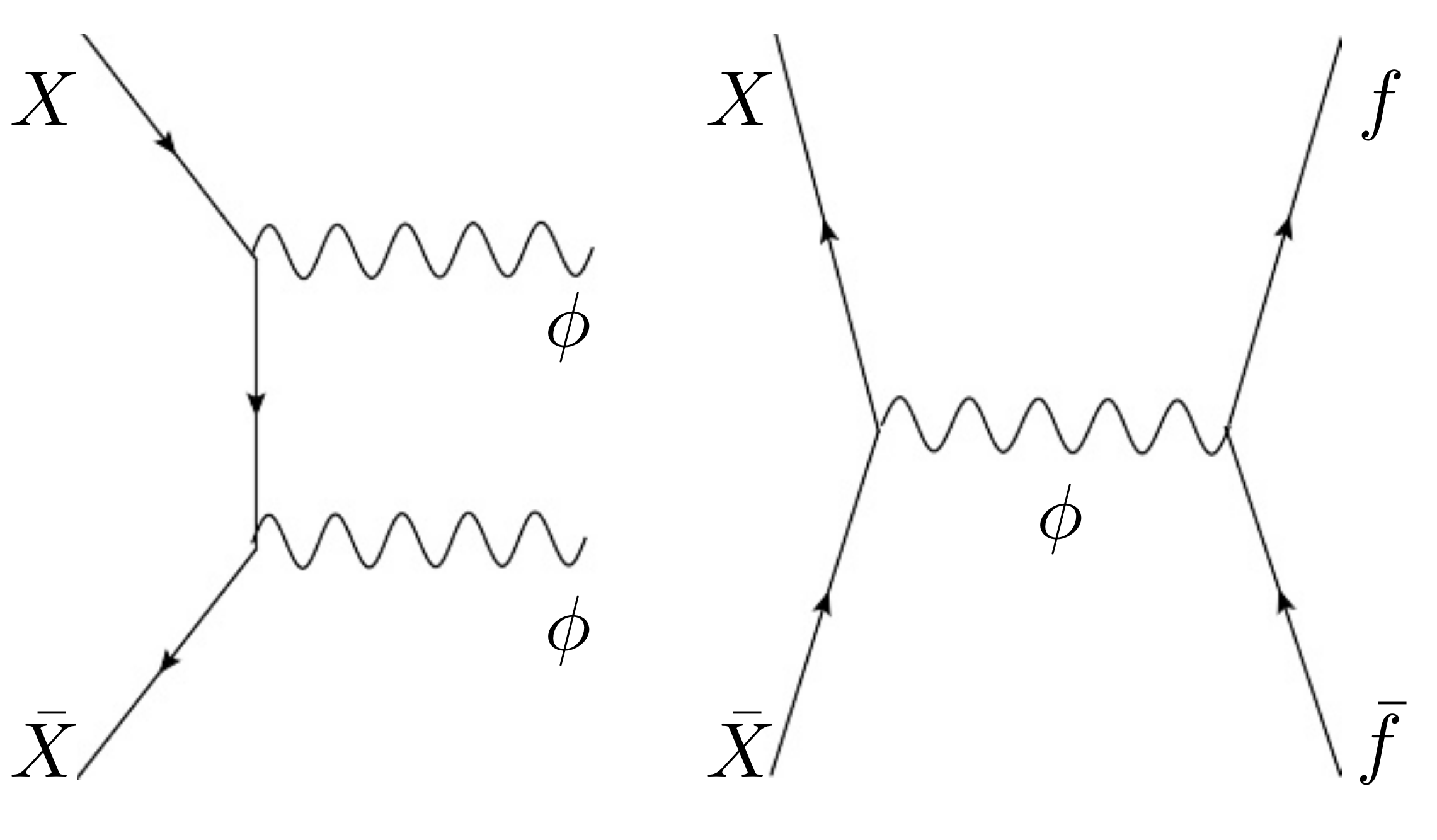}
\caption{Typical processes for removing the thermal symmetric (in $X$ and $\bar X$) abundance.  $X-\bar X$ can annihilate to light force mediators $\phi$ (whether they be scalar or vector), or through a heavier state directly to SM $f \bar f$.}
\label{Diagrams}
\end{center}
\end{figure}

With an existing asymmetry in both the DM and visible sectors (whether simultaneously generated at baryo- or darko-genesis, or through a transfer mechanism at a later time), a small asymmetry floats on top of a large thermal abundance.  This thermal abundance must be efficiently removed through some mechanism.  There are predominantly two ways to do this: through direct annihilation to light mediators, as in the left panel of Fig.~\ref{Diagrams}, or through mediators that are heavier than the DM, as in the right panel. 

When the DM carries a substantial asymmetry, the usual freeze-out calculation for the DM relic abundance is modified.  Annihilations of $X \bar{X}$ remove most of the thermal symmetric component of DM, leaving mostly the asymmetric component.  How much of the symmetric component remains depends on the details of the freeze-out calculation when the non-zero asymmetry is included.  We review this calculation, as layed out in \cite{Graesser:2011wi,Lin:2011gj,Iminniyaz:2011yp}, before moving on to discussing models for annihilating away the symmetric component.  We will discuss only $D$ preserving annihilation processes; the effect of $D$ violating wash-out processes on the relic abundance is considered in \cite{Ellwanger:2012yg}.

We describe the relic density calculation in terms of the relic asymmetry, $r_\infty
= \Omega_{\bar X}/\Omega_X$.  This controls the size of indirect
signals from DM annihilation. $r_\infty$ is related to the
absolute relic densities by
\begin{align}
  \Omega_X = \frac{1}{1-r_\infty} \frac{\eta_X m_X s_0}{\rho_c}, \ \
  \Omega_{\bar X} = \frac{r_\infty}{1-r_\infty} \frac{\eta_X m_X s_0}{\rho_c}, \ \
  \label{eq:omegar_relation}
\end{align}
and the total CDM relic density is $\Omega_c = \Omega_X +
\Omega_{\bar{X}}$.
We take the standard definitions $x=m_X/T$ and $Y_\pm = n_\pm
/s$, where $s=(2\pi^2/45)h_{eff}(T)T^3 $ is the entropy density and
$h_{eff}(T)$ is the effective number of degrees of freedom for the
entropy density. We write the annihilation cross section as
$\sigv=\sigma_0x^{-n}$, with $n=0$ and $n=1$ for $s$-wave and $p$-wave
annihilation processes respectively, so that the Boltzmann equation gives
\begin{align}
  \frac{dY_\pm}{dx} &= - \frac{\lambda}{x^{n+2}} \sqrt{g_*} \left( Y_+ Y_-  - (Y^{eq})^2 \right)
  \label{eq:boltzmannY}
\end{align}
where $\lambda\equiv 0.264 \mpl m_X\sigma_0$ and
$Y^{eq}\simeq0.145(g/h_{eff})x^{3/2}e^{-x}\equiv a x^{3/2}e^{-x}$.   The DM asymmetry is 
$\eta_X = Y_+ -Y_-$.

This equation can be solved analytically at late times when
$(Y^{eq})^2$ becomes negligible, to obtain
\begin{align}
  Y_\pm (\infty)&\simeq \frac{\pm \eta_X}{1-\left[1 \mp\eta_X/{Y_\pm}(x_f)\right]e^{\mp \eta_X\lambda \sqrt{g_*} x^{-n-1}_f/(n+1)}}.
  \label{eq:boltzmann_latetime}
\end{align}
The freezeout temperature $x_f=m_X/T_f$ is
derived in~\cite{Graesser:2011wi}:
\begin{equation}
  x_f\simeq\ln\left[(n+1)\sqrt{g_*}a\lambda\right]+\frac{1}{2}\ln\frac{\ln^2\left[(n+1)\sqrt{g_*}a\lambda\right]}{{\ln^{2n+4}\left[(n+1)\sqrt{g_*}a\lambda\right]-(\sqrt{g_*})^2\left[(n+1)\lambda\eta_X/2\right]^2}}.
\end{equation}
Using $Y_\pm(\infty)$ given in \eq{eq:boltzmann_latetime}, we obtain:
\begin{equation}
  r_\infty \equiv \frac{Y_-}{Y_+}(\infty) \simeq \frac{Y_-(x_f)}{Y_+(x_f)} \exp \left( \frac{ -\eta_X\lambda \sqrt{g_*}}{ x_f^{n+1} (n+1) } \right).
  \label{eq:rinfty}
\end{equation}

As we will see in Sec.~\ref{sec:cosmoconstraints}, an annihilation cross-section $\sigma v \gtrsim 10^{-25} \mbox{ cm}^3/\mbox{s}$ is needed to evade CMB constraints if the annihilation is $s$-wave.  Obtaining such large cross-sections is challenging consistent with current constraints, if the annihilation goes through a heavy mediator.

\subsection{Heavy Mediators}

Ref.~\cite{Kaplan:2009ag} found that it is difficult to annihilate the symmetric component sufficiently through weak scale mediators (consistent with collider constraints) if the DM is $\sim 10 \mbox{ GeV}$.  Parameterically, the annihilation cross-section through a heavy mediator $M$ to a pair of fermions is
\begin{equation}
\sigma_{ann} v = \frac{g_d^2 g_f^2 m_X^2}{\pi m_M^4} \simeq 10^{-26} \mbox{ cm}^3/\mbox{s} \left(\frac{g_d g_f}{0.25}\right)^2 \left(\frac{10 \mbox {GeV}}{m_X}\right)^2 \left(\frac{200 \mbox{ GeV}}{m_M}\right)^4,
\end{equation} 
where $g_d,~g_f$ are the mediator couplings to the DM and fermions, respectively.  Given collider constraints on objects around the weak scale, one can immediately see that it may be difficult to obtain a sufficiently large annihilation cross-section.

More generally, monojet constraints from the Tevatron and LHC make this statement more robust \cite{Buckley:2011kk,MarchRussell:2012hi}.  The basic idea of the standard mono-jet searches is to make use of a jet radiated off the initial state to constrain higher dimension operators through which the DM might annihilate.  Possible operators are, for example, of the form (an exhaustive list is given in \cite{Goodman:2010ku})
\begin{eqnarray}
{\cal O}_1&  = & \frac{1}{\Lambda^2}\bar{X} X \bar{f} f \\ 
{\cal O}_2 & = & \frac{1}{\Lambda^2}\bar{X} \gamma^\mu X \bar{f} \gamma_\mu f .
\label{AnnOps}
\end{eqnarray}

\begin{figure}
\begin{center}
\includegraphics[width=0.6\textwidth]{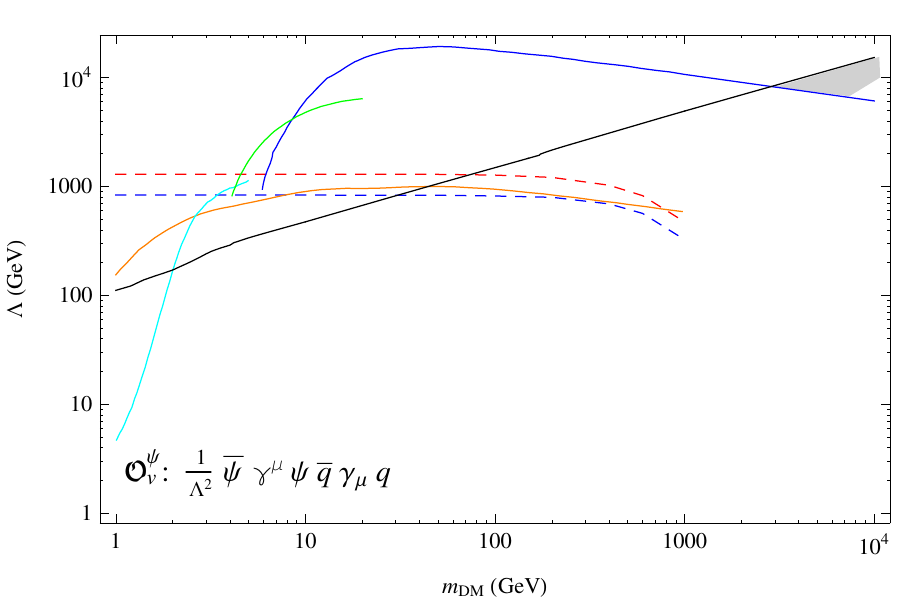}
\caption{Constraints on the scale $\Lambda$ of a higher dimension operator through which the DM can annihilate to SM quarks.  The black curve corresponds to the minimum annihilation cross-section necessary to remove the symmetric component to 1\%.  The solid curves represent constraints from various direct detection experiments, while the dashed curves show constraints from ATLAS and CMS monojet searches.  Only the grey shaded region is consistent with sufficiently annihilating the symmetric abundance and satisfying all the constraints.  In this region the DM mass is well outside the natural ADM window of 1-10 GeV.  See \cite{MarchRussell:2012hi} for more details of the analysis.}
\label{MonojetAndDD}
\end{center}
\end{figure}

Monojet plus missing energy measurements at the LHC combined with direct detection rule out scales $\Lambda$ such that DM annihilation is sufficient through these operators to deplete the symmetric component to $1\%$ of its asymmetric abundance, {\em if the mediator particle is at the weak scale or heavier}.  This is shown in Fig.~\ref{MonojetAndDD}.  The lines in the figure correspond to constraints on the scale $\Lambda$ from monojet and direct detection searches, while the grey line indicates the upper bound on $\Lambda$ consistent with the DM density.  One can see that there is only a consistent region (shaded grey) at large DM mass, which is above the natural ADM window.

As a result, direct annihilation to light states in the DM sector is preferred over heavy states for depleting the symmetric component.  This is analogous to the visible sector, where the presence of the electromagnetic force efficiently removes the excess of positrons and anti-baryons to leave only the relic baryon and lepton asymmetry.

We should note, however, that these strong constraints on annihilation through a mediator heavier than $2 m_X$ from colliders are only robust if the mediator particle is at the weak scale or heavier.  The constraints are much weakened for a lighter particle, and sufficient annihilation may be obtained, especially when the mediator particle is on resonance, $m_M = 2 m_X$ \cite{Feng:2012jn}.

\subsection{Dark Forces}

By extension, the most natural way for DM to annihilate is to dark states \cite{Cohen:2010kn,Lin:2011gj,Blennow:2012de,Bhattacherjee:2013jca}.  These states may be stable, which will remain as dark radiation today \cite{Blennow:2012de}.  Since cosmological data is (roughly) consistent with three neutrino species \cite{Ade:2013zuv}, stringent constraints must be met by the dark sector.  On the other hand, DM may annihilate to dark force mediators, whether they be scalars or vectors, as shown in Fig.~\ref{Diagrams}, that subsequently decay to SM particles.    The annihilation cross-section to a pair of scalars or vectors is
\begin{equation}
\langle \sigma v \rangle_V \simeq \frac{\pi \alpha_X^2}{m_X^2}\sqrt{1-\frac{m_V^2}{m_X^2}},~~~~~\langle \sigma v \rangle_S \simeq \frac{9}{4} \frac{\pi \alpha_X^2}{m_X^2}\frac{T}{m_X}\sqrt{1-\frac{m_\phi^2}{m_X^2}},
\label{annihilateThermalAbundance}
\end{equation}
where $\alpha_X = g_X^2/4\pi$.
If $m_M > m_X$ (but not much heavier), then annihilation can also proceed through an $s$-channel exchange of the mediator
\begin{equation}
\langle \sigma v \rangle_V \simeq 4 \alpha_X g_f^2 m_X^2 N_f^c/m_\phi^4,~~~~~\langle \sigma v \rangle_S \simeq \alpha_X g_f^2 m_X^2 N_f^c/2 m_\phi^4 x_f.
\end{equation}

For ADM, the CMB places a potentially important {\em lower} bound \cite{Lin:2011gj} on the annihilation cross-section, to sufficiently remove the symmetric component such that ionizing radiation is not injected into the spectrum at late time.  For nearly 100\% ionizing efficiency ({\em i.e.} for annihilation to $\mu^+ \mu^-$ or $e^+ e^-$), the lower bound for 10 GeV DM is approximately $10^{-25} \mbox{ cm}^3/\mbox{s}$.  If the ionizing efficiency is lower ({\em e.g.} for annihilation to pairs of $\tau$'s), the constraints are weaker.  We review these constraints in more detail below in Sec.~\ref{sec:cosmoconstraints}.

\section{ADM Hidden Sectors}
\label{sec:hidden sectors}

While the DM sector is conventionally envisioned to be simple, consisting of the DM particle only, this need not necessarily be the case.  The structure of the DM sector may be rich, with multiple DM states as well as dark forces.  In fact, as we saw in the previous section, light particles in the hidden sector may be important for annihilating the symmetric relic abundance of DM, leaving only the asymmetric component.  As we will see in this section, dark sectors with structure may also set the mass scale in the DM sector, as well as give rise to self-interacting DM.  In this section, we review hidden sector models, as well as cosmological implications of self-interacting DM on the large scale structure of the universe.  Before we proceed, we note that while dark hidden sectors provide useful functions in the context of ADM, they are not unique to ADM.

\subsection{Hidden Sector Models}

More predictive models of ADM arise when taking more ambitious goals for the DM sector than simply annihilating away the symmetric abundance of DM.  For example, the DM sector may have a structure such that it mirrors the visible sector, naturally explaining why the dark and visible mass scales are in the same range \cite{Hodges:1993yb,Mohapatra:2000rk,Foot:2004pq}.  The dark sector may have a dark QCD, as in Hidden Valley models \cite{Strassler:2006im}, with the mass set by the confinement scale, similar to the the proton \cite{Bai:2013xga}.

In some ways the most predictive models of ADM, however, are those where the mass scale of the DM sector is connected to the strength of supersymmetry breaking in the hidden sector.  If supersymmetry breaking is weaker in the DM sector than in the visible sector, the DM may naturally be lighter than the weak scale. 

\begin{figure}
\begin{center}
\includegraphics[width=0.45\textwidth]{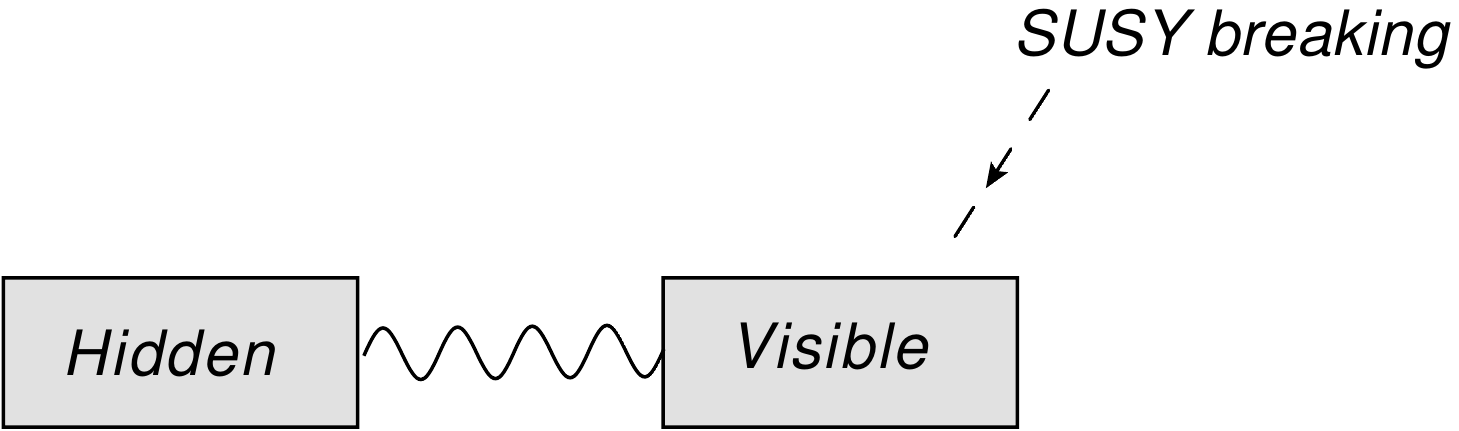}
\includegraphics[width=0.45\textwidth]{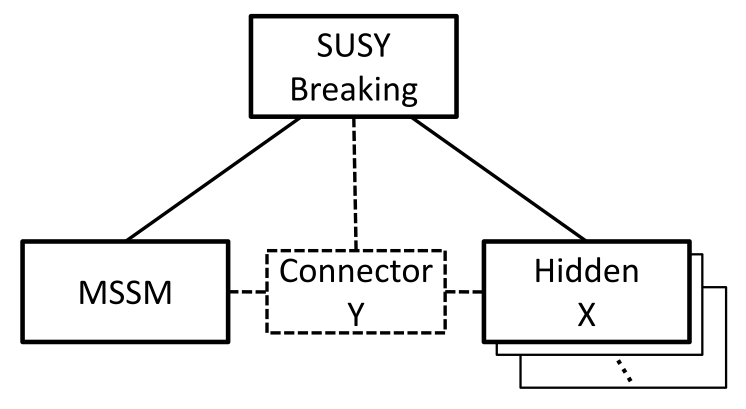}
\caption{Schematics of two means of communicating supersymmetry breaking to the DM sector.  In the left figure, SUSY breaking is communicated to the visible (SM) sector, which is then weakly communicated to the hidden sector via matter or gauge interactions, as in \cite{Hooper:2008im}.  In the right figure \cite{Kumar:2009bw} SUSY breaking is directly communicated to both hidden and visible sectors via gauge interactions, but the hidden sector is lighter because SUSY breaking is communicated more weakly to the DM than the baryons.}
\label{hiddenSector}
\end{center}
\end{figure}

There are two ways that supersymmetry breaking can set the mass scale in the DM sector.  These two scenarios are shown schematically in Fig.~\ref{hiddenSector}.  First, the dark sector may couple to the visible sector, communicating the visible sector supersymmetry breaking to the dark sector \cite{Hooper:2008im,ArkaniHamed:2008qn,Zurek:2008qg,Baumgart:2009tn,Morrissey:2009ur}, as shown in the left panel of Fig.~\ref{hiddenSector}.  The supersymmetry breaking effects in the DM sector are suppressed by a loop factor in comparison to the weak scale, naturally explaining why the DM may have a mass around the GeV scale.  As a simple example, take the interaction of a singlet $S$ with the supersymmetric SM, $\lambda S H_u H_d$ \cite{Zurek:2008qg,Morrissey:2009ur}.  The coupling of $S$ to $H_u,~H_d$ generates a supersymmetry breaking mass term for $S$ of the size
\begin{equation}
\delta m_S^2 \simeq - \frac{\lambda^2}{8 \pi^2}(m_{H_u}^2 + m_{H_d}^2),
\label{darkloop}
\end{equation} 
where $m_{H_u}^2,~m_{H_d}^2$ are the soft masses for the up and down type Higgs doublets.\footnote{Since this mass squared term is negative, an additional term, such as a quartic, must stabilize the field.  Various mechanisms can easily do this.}  If $\lambda$ is not too far from 1, GeV scale masses in the hidden sector are naturally generated via the loop suppression factor.  Supersymmetry breaking can also be communicated to the dark sector via gauge kinetic mixing of a dark gauged $U(1)_X$ with hypercharge \cite{Hooper:2008im,ArkaniHamed:2008qn,Baumgart:2009tn,Morrissey:2009ur}, 
\begin{equation}
{\cal L}_{mix} = \frac{\epsilon}{2} F^{\mu \nu} F'_{\mu \nu},
\end{equation}  
where $F'_{\mu \nu}$ is the $U(1)_X$ field strength and $F_{\mu \nu}$ is the hypercharge field strength.  Supersymmetry breaking in the visible sector is transferred to the dark sector via the hypercharge $D$-term \cite{Baumgart:2009tn}, $\xi_Y = -\frac{g_Y}{ 2} c_{2\beta} v^2$, where $v$ is the Higgs vev, $g_Y$ is the hypercharge gauge coupling, and $\tan \beta$ is the usual ratio of up to down type Higgs vevs.  This $D$-term is communicated to the dark sector with strength $\xi = \frac{\epsilon}{g_X} \xi_Y$, which is naturally $\sim \mbox{ GeV}^2$ if the kinetic mixing parameter $\epsilon$ is ${\cal O}(10^{-3})$ \cite{ArkaniHamed:2008qn}, as is typical in kinetic mixing models \cite{Holdom:1985ag}. This can induce a GeV scale vev for one of the scalars in the DM sector, setting the natural mass scale for the other particles in the theory.  For example, if we have two oppositely charged $U(1)_X$ dark scalars, $H,~H'$, we have the scalar potential from the $D$-terms which is
\begin{equation}
V_D = \frac{g_X^2}{2}(|H|^2-|H'|^2-\xi)^2,
\label{kineticmixDterm}
\end{equation}
so that $\langle|H|\rangle = \sqrt{\xi}$.

The second way to set the mass scale in the dark sector via supersymmetry breaking is by direct coupling to the supersymmetry breaking sector itself, and not via the supersymmetric SM.  This is shown in the right panel of Fig.~\ref{hiddenSector}.  For example, within gauge mediated supersymmetry breaking models, the scale for soft scalar masses in the MSSM is
\begin{equation}
m_{\phi^k}^2 = 2 \sum_{i=1}^3 \frac{\alpha_i^2}{16 \pi^2} C_i(k) \frac{F^2}{M^2},
\end{equation}
where $F$ and $M$ are the SUSY-breaking $F$-term and vev of the messenger field, respectively (see \cite{Martin:1997ns} for a review), $\alpha_i\equiv g_i^2/(4\pi)$ are the SM gauge couplings, and $C_i$ are the quadratic Casimir invariants.  If the messengers are likewise charged under a dark gauge group, $U(1)_X$, supersymmetry breaking is communicated to the DM sector with strength \cite{Hooper:2008im,Kumar:2009bw}
\begin{equation}
m_{dark}^2 = 2 \frac{\alpha_X^2}{16 \pi^2} \frac{F^2}{M^2},
\label{darkgauged}
\end{equation}
where $\alpha_X$ is the $U(1)_X$ gauge coupling.  Thus, if $g_X \ll g_i$, then the natural mass scale in the hidden sector is much lower than in the visible sector.

We now discuss a concrete model in the context of ADM, using the tools we just discussed \cite{Hooper:2008im,Zurek:2008qg,Baumgart:2009tn,Morrissey:2009ur}.   See \cite{Cohen:2010kn} for more details of this model, though we outline the general features and note that many of these features will be common to any theory that sets the DM mass scale via a supersymmetric hidden sector (see for example \cite{Blinov:2012hq}).  The ADM sector is extended to include a dark Higgs, $H'$, charged under a dark gauged $U(1)_X$, and the matter content is extended to include the sterile $X$ as well as a matter partner $T$ that is charged under the $U(1)_X$ such that a mass term \cite{Cohen:2010kn}
\beq
W = \lambda X T H'
\eeq 
arises in the hidden sector.  The asymmetry is communicated between sectors via the usual ADM operator (Eq.~\ref{highdimops})
\beq
W = X^n {\cal O}_{B-L}.
\eeq
In addition, there is the hypercharge-$U(1)_X$ kinetic mixing term connecting the dark and visible sectors that, as we discussed above, induces a vev for $H'$ via Eq.~\ref{kineticmixDterm}.
For $\epsilon \sim 10^{-3}$ and the gauge coupling $g_X \sim 1$, this vev is around 10 GeV, inducing a 10 GeV mass for the $X-T$ Dirac fermion state if $\lambda$ is ${\cal O}(1)$.  The scalar super-partners of $X$ and $T$ will obtain the same mass, up to important SUSY breaking corrections. Two loop gauge-mediated graphs via Eq.~\ref{darkgauged} will push the mass of the $T$ and $H'$ scalars up relative to the fermions.  $X$ is a singlet, so it does not receive these contributions.  It does, however, receive a 1-loop contribution from the $T$ and $H'$ scalar soft SUSY masses that pushes the $X$ scalar {\em down} relative to the fermion, via Eq.~\ref{darkloop} (with the right-hand side of this equation having the $T$ and $H'$ soft masses instead of $H_u$ and $H_d$).  Thus the scalar $X$ is the DM in this model.  

The dark photon and its superpartner are lighter than the DM if $g_X \ll \lambda$.  The DM can efficiently annihilate to the dark gauge boson that subsequently decays back to the standard model, via the same mechanism discussed in Sec.~\ref{sec:annihilation}.  The dark photino decays to the gravitino and a photon, which occurs quickly enough to evade big bang nucleosynthesis constraints if the supersymmetry breaking scale is low \cite{Cohen:2010kn}.

The direct detection signal also proceeds via the kinetic mixing with $U(1)_X$.  While $X$ is a singlet under $U(1)_X$, a 1-loop graph via $T$ and $H'$ can induce a coupling to the dark photon, $\gamma_d$:
\begin{eqnarray}
& \frac{\lambda^2 g_X}{16 \pi^2}\left(\frac{4 g_X^4 - \lambda^4 + 4 \lambda^2 g_X^2 \log \left(\frac{\lambda^2}{2 g_X^2}\right)}{2(2 g_X^2 - \lambda^2)^2}\right) X^\dagger \partial_\mu X \gamma_d^\mu \\ \nonumber
& \equiv g_X q_{eff} X^\dagger \partial_\mu X \gamma_d^\mu.
\end{eqnarray}
Scattering of the DM off the proton (or any charged particle) can then proceed via kinetic mixing of the dark photon with the SM photon.  The cross-section is
\beq
\sigma_{p,X} = \frac{4}{\pi}\frac{g_W^4 c_W^4 \mu_{X,p}^2}{c_{2 \beta}^2 m_W^4} q_{eff}^2.
\eeq
In this expression, we have canceled $g_X$ dependence that enters into the dark photon mass as well as the coupling, so that the final result is independent of $g_X$:
\beq
\sigma_{p,X} \simeq 9.2 \times 10^{-42} \mbox{ cm}^2 \lambda^4.
\eeq
We will discuss in Sec.~\ref{sec:directindirectdetection} direct and indirect detection constraints on ADM, but we note here that this scattering cross-section sits in the range needed for the CDMS \cite{Agnese:2013rvf} and CoGeNT \cite{Aalseth:2010vx,Aalseth:2011wp,Aalseth:2012if} anomalies, and is reachable with many of the next generation direct detection experiments.  

These hidden sectors give rise to a broad range of signatures, notably in experiments with an intense electron or proton beam.  This has given rise to a rich experimental program ({\em e.g.} \cite{Bjorken:2009mm,Merkel:2011ze,Abrahamyan:2011gv}; see \cite{Izaguirre:2013uxa,Essig:2013vha} for a more complete set of references).  These experiments are typically direct probes of the mediating particle (the dark photon in the example above), and are complimentary probes to direct detection.

The dark forces mediate scattering not only with nuclei, but also with the DM itself.  This self-scattering has potentially important cosmological implications, which we discuss next.

\subsection{Self-interacting Dark Matter Through Dark Forces}

In the standard WIMP paradigm, DM carries little or no self-interactions.  On the other hand, with the introduction of light dark forces, for example to annihilate the thermal relic abundance, as in Sec.~\ref{sec:annihilation}, or to set the mass scale in the DM sector, as we just described above, self-interactions appear naturally.  The scattering cross-section through a force $\phi$ with dark structure constant $\alpha_X$ coupling to the DM is
\beq
\sigma_X \approx 5 \times 10^{-23} \mbox{ cm}^2\left(\frac{\alpha_X}{0.01}\right)^2\left(\frac{m_X}{10 \mbox{ GeV}}\right)^2\left(\frac{10 \mbox{ MeV}}{m_\phi}\right)^4,
\label{scattering}
\eeq 
so that self-scattering can be important even for moderate self-coupling.

Self-interactions can have important cosmological implications.  Indeed, there have long been questions about whether DM self-interactions can help alleviate some of the apparent inconsistencies between the standard collisionless Cold Dark Matter (CCDM) paradigm and the observations of structure.  The most notable problems with the CCDM paradigm to date are (see Sec.~II of \cite{Tulin:2013teo} for a more detailed summary and references):
\begin{itemize}
\item Core-vs-cusp problem.  The central densities of both dwarf galaxies and cluster of galaxies are observed to be lower than predicted by CCDM \cite{Wechsler:2001cs,Dubinski:1991bm}, displaying a core rather than a cusp \cite{Flores:1994gz,Simon:2004sr}.  
\item Missing satellites problem.  The number of observed satellite dwarf galaxies in the Milky Way is about an order of magnitude lower than predicted by simulations of CCDM \cite{Klypin:1999uc,Moore:1999nt,Bullock:2010uy}.  The Sloan Digital Sky Survey, however, has discovered many faint dwarf galaxies such that it is apparent that as many as 20 times more dwarf galaxies could remain undiscovered due to faintness or sky coverage \cite{Bullock:2009gv,Bullock:2010uy}.  Thus the missing satellites problem is by many no longer viewed as a problem for CCDM.
\item Too big to fail problem.  The brightest dwarf spheroidal galaxies in the Milky Way are expected to be hosted by the most massive DM sub halos in the galaxy.  The sub halos predicted by the simulations to host the dwarf spheroidal galaxies are not, however, observed to exist: simulations predict ${\cal O}(10)$ sub halos with maximum circular velocity $v_{max} > 30 \mbox{ km/s}$, while the observed dwarf spheroidal galaxies are all hosted in halos with $v_{max} < 25 \mbox{ km/s}$  \cite{BoylanKolchin:2011de,BoylanKolchin:2011dk,Walker:2012td,Strigari:2007ma}.  This can be re-stated in terms of the core-vs-cusp problem: the hosts of the brightest dwarf spheroidal galaxies have lower central densities than predicted by CCDM.
\end{itemize}

The paradigm of self-interacting DM (SIDM)  \cite{Spergel:1999mh} was introduced to solve some of these problems with DM self-interactions.  As DM virializes, self-scattering can transfer heat from the outer part of the halo to the inner part of the halo, flattening the density profile of the inner part of the halo, as required to solve both the core-vs-cusp problem and the too big to fail problem.  Simulations have shown that the needed cross-section is on the order of $\sigma/m_X \simeq 10^{-24} \mbox{ cm}^2/\mbox{ GeV}$ \cite{Dave:2000ar,Yoshida:2000uw,Gnedin:2000ea,Rocha:2012jg,Zavala:2012us}.  Further investigation of SIDM also showed, however, that gravitational instabilities can result at large velocity dispersion ({\em i.e.} in large objects, such as clusters of galaxies) if the scattering cross-section has no velocity dependence, though some recent studies have called this conclusion into question \cite{Rocha:2012jg,Peter:2012jh}. 

Since self-interactions of the right size appear naturally in models of ADM, as in Eq.~\ref{scattering}, it is natural to consider the range of dynamics that can appear, since the size of the scattering cross-section directly impacts the halo shape and profile for objects of different mass ({\em i.e.} velocity dispersion).  Mirror dark matter was posited as an asymmetric mirror sector of the visible one, with DM self-interactions, largely motivated by solving these structure problems \cite{Berezhiani:1995yi,Mohapatra:2000rk,Mohapatra:2000qx,Berezhiani:2000gw,Mohapatra:2001sx,Ignatiev:2003js,Foot:2003jt,Berezhiani:2003wj,Foot:2004wz}.  The simulations have to this point  mostly assumed a constant scattering cross-section (with recent exceptions \cite{Vogelsberger:2012ku,Zavala:2012us}), but when light dark forces are introduced, a wide variety of dynamics can ensue \cite{Loeb:2010gj,Tulin:2012wi,Tulin:2013teo}.  The scattering cross-sections become velocity dependent in the Coulomb regime (when $m_X v \gtrsim m_\phi$), where the scattering cross-section scales as a Rutherford cross-section,
\beq
\sigma_{\mbox{clas}} = \frac{16 \pi \alpha_X^2}{m_X^2 v^4}\log(1+m_X^2 v^2/m_\phi^2) .
\eeq
On the other hand, in the non-perturbative regime ($\alpha_X m_X/m_\phi \gg 1$), resonant effects can become important \cite{Tulin:2012wi,Tulin:2013teo}, and this impacts structure formation.  Depending on the parameters of the the theory, the DM self-interaction cross-section can vary widely for different mass objects.  Thus scattering may be unimportant for DM halos in clusters of galaxies, but crucial for the formation of cores in dwarf spheroidal galaxies.

While DM self-interactions arise naturally within ADM models, they are not a smoking gun signature for ADM.  We next consider the evolution of the primordial DM asymmetry, and its impacts on smoking gun signatures for ADM.

\section{Asymmetry Wash-Out Via Oscillations}
\label{sec:oscillations}

The phenomenology of ADM in the universe today depends in large part on whether the DM retains its primordial asymmetry until today.  Before discussing astrophysical constraints on ADM in Sec.~\ref{sec:cosmoconstraints} we thus consider under what circumstances the asymmetry will be retained versus washed away.

If DM has not only the dark number $D$ preserving mass term, $m_X \bar X X$ (or $m_X^2 |X|^2$ for scalars), but also an $D$-violating mass term $m_M \bar X^c X$ (or $m_M^2 X^2$ for scalars), then the DM asymmetry may be washed out via particle-anti-particle oscillations \cite{Cohen:2009fz,Buckley:2011ye,Cirelli:2011ac,Tulin:2012re}.  Here $X^c = - i \gamma^2 X^*$ for fermions and $X^c = X^\dagger$ for scalars. The oscillations occur when the Dirac or complex scalar state is split via the Majorana-type mass term into two real states with mass $m \pm \delta m$, where
\beq
\delta m \equiv  \left\{ 
\begin{tabular}{cl}
$m_M$ & fermionic X case \\
$\frac{m_M^2}{2 m_X}$ & scalar X case
\end{tabular}
\right. .
\eeq
This mass splitting causes $X$ and $X^c$ to mix, in a way that is similar to neutrino flavor mixing, but with DM number violation replacing neutrino flavor violation.  This mixing leads to $X$ flavor oscillations into $X^c$.    Naively, the asymmetry should be washed out when the Hubble expansion rate drops below the mass splitting, $H \lesssim \delta m$.  If we require that the asymmetry be erased only at temperatures below which the thermal symmetric component has already annihilated away ($T \lesssim 1 \mbox{ GeV}$ for 10 GeV DM), this naively implies $\delta m \lesssim 10^{-19} \mbox{ GeV}$.  Such a small mass splitting may indicate the presence of a gauged symmetry to protect the global $U(1)_D$, as is often invoked for, {\it e.g.}, $B-L$.   Though wash-out may occur once $H \lesssim \delta m$, it is crucial to take into account scattering and annihilation rates, and their sensitivity to DM flavor.  These effects change the onset of oscillations in a qualitative way. 
We follow the analysis of \cite{Tulin:2012re}, which fully accounted for all these effects, including scattering and the sensitivity to the type of interaction (scalar or vector), in the evolution.

The evolution of the particle-anti-particle asymmetry can be seen explicitly by writing out the mass matrices and considering the Boltzmann equation in the absence of collisions.  In the ``flavor'' basis $(X,X^C)$, the mass matrices for the fermions ${\cal M}_f$ and scalars ${\cal M}^2_S$ are
\beq
{\cal M}_f = \left(
\begin{tabular}{cc}
$m_X$ & $m_M$ \\
$m_M$ & $m_X$
\end{tabular}
\right) ~~~~~~~~~~~
{\cal M}^2_S = \left(
\begin{tabular}{cc}
$m_X^2$ & $m_M^2$ \\
$m_M^2$ & $m_X^2$
\end{tabular}
\right), 
\eeq
so that the non-interacting Boltzmann equation is
\beq
\frac{\partial {\cal F}_k}{\partial t} - H k \frac{\partial {\cal F}_k}{\partial k} = - i\left[{\cal H}_k,{\cal F}_k \right].
\eeq
Expanding the free Hamiltonian ${\cal H}_k$, we have
\beq
{\cal H}_k = \sqrt{k^2 + {\cal M}^2} = \omega_k + \frac{m_X \delta m}{\omega_k}\left(\begin{tabular}{cc} 0 & 1 \\ 1 & 0 \end{tabular}\right),
\eeq
where $\omega_k = \sqrt{k^2 + m_X^2}$.  As a result, when the DM is cold so that $m_X \sim \omega_k$, wash-out of the DM asymmetry can occur when $\delta m \sim H$.
Thus the Majorana mass must be very small to avoid any effects from flavor mixings.

Wash-out does not necessarily occur, however, once oscillations commence.  Coherence of the flavor wave function, as well as the flavor sensitivity of the interactions, is extremely important. The basic results can be summarized as follows:
\begin{itemize}
\item Rapid scatterings prevent oscillations from occurring.  This is a quantum Zeno (or ``watched pot'') effect.  Thus no wash-out can occur until scatterings have decoupled.
\item When oscillations commence, washout does not necessarily commence.  If flavor coherence is maintained, and the interactions are flavor sensitive ({\em e.g.} for vector particles mediating DM annihilation to fermion pairs), no wash-out can occur.  Coherence is maintained in the absence of scatterings.
\end{itemize}

We outline the important aspects of the calculation and refer the reader to \cite{Tulin:2012re} for details. The Boltzmann equation for the system is
\beq
Y'(x) = -\frac{i}{H x}\left[{\cal H}_0,Y\right] - \frac{\Gamma_\pm}{2 H x}\left[ O_\pm,\left[O_\pm,Y\right]\right]-\frac{s \langle \sigma v \rangle_\pm}{H x} \left(\frac{1}{2}\left\{Y,O_\pm \bar{Y} O_\pm \right\} - Y_{eq}^2 \right)
\label{Boltzmann}
\eeq
where $x \equiv m_X / T$, $Y_{eq} \equiv n_{eq}/s$, and ${\cal H}_0$ is defined according to
\beq
{\cal H}_0 \equiv \left(\begin{tabular}{cc} $m_X$ & $\delta m$ \\ $\delta m$ & $m_X$ \end{tabular}\right).
\eeq 
Here the $O_\pm$ encode the flavor information of the scatterings
\beq
{\cal L}_{\rm int} = \bar{\Psi} \Gamma^a O_\pm \Psi \bar{f} \Gamma_a f,~~~~O_\pm \equiv \left( \begin{array}{cc} 1 & 0 \\ 0 & \pm 1 \end{array} \right).
\eeq
Scalar, pseudo scalar, and axial-vector interactions are flavor-blind ($O_+$), while vector and tensor interactions are flavor sensitive ($O_-$).

We can see the effects on the annihilation and scattering terms.  In the annihilation terms
\begin{subequations}
\begin{align}
&\textrm{flavor-blind:}  & \frac{1}{2} \big\{ Y, \, O_+ \bar{Y} O_+ \big\} \;=& \;\left( \begin{array}{cc} Y_{11} Y_{22} + Y_{12} Y_{21} & Y_{11} Y_{12} + Y_{12} Y_{22} \\  Y_{21} Y_{11} + Y_{22} Y_{21} & Y_{11} Y_{22} + Y_{12} Y_{21} \end{array} \right)\\
&\textrm{flavor-sensitive:}  & \frac{1}{2} \big\{ Y, \, O_- \bar{Y} O_- \big\} \;=&\; \left( \begin{array}{cc} Y_{11} Y_{22} - Y_{12} Y_{21} & 0 \\  0 & Y_{11} Y_{22} - Y_{12} Y_{21} \end{array} \right) \; .
\label{annihilations}
\end{align}
\end{subequations}
Note that in the absence of coherence ($Y_{12},~Y_{21}\rightarrow 0$), flavor sensitive and blind interactions give the same result for annihilation.  In the presence of coherence, however, flavor sensitivity matters. Flavor sensitive interactions, as can be seen in Eq.~\ref{annihilations}, couple to the determinant of Y.  On the other hand, $\mbox{det}([{\cal H}_0,Y]) = 0$, which implies that coherent flavor sensitive interactions do not lead to annihilation.  Annihilations via flavor-blind interactions, on the other hand, proceed regardless of whether coherence is present.  

\begin{figure}
\begin{center}
\includegraphics[width=0.95\textwidth]{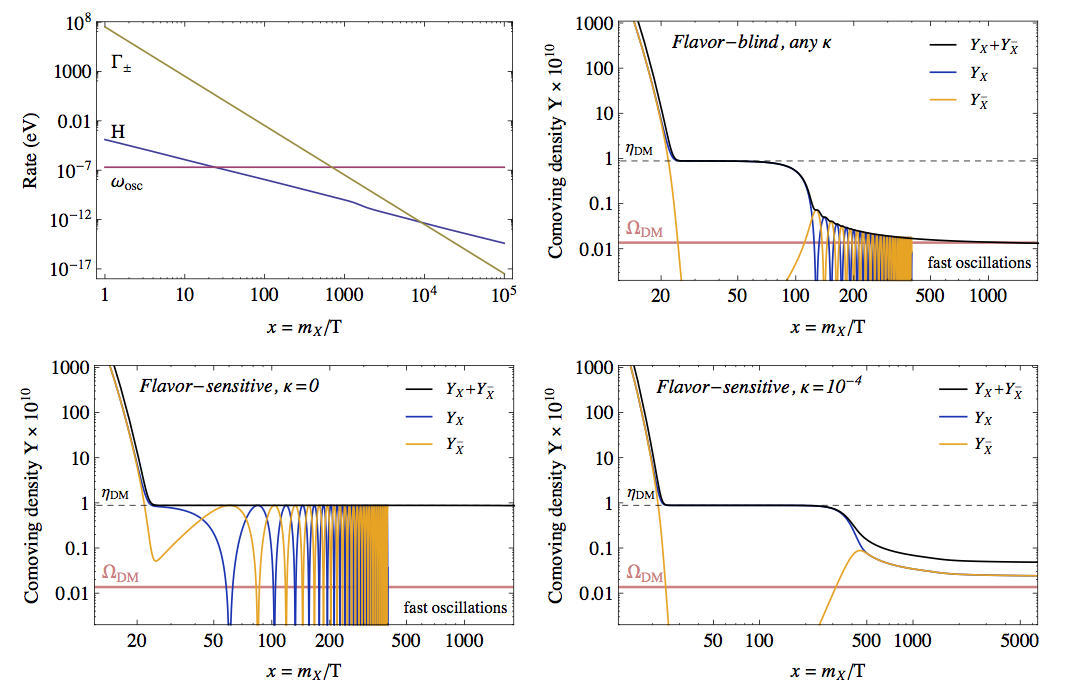}
\caption{The effects of scattering and the ``flavor'' sensitivity of the annihilation process.  The first panel shows the relative rates of scattering, oscillation and Hubble expansion $H$.  Once $H$ drops below $\omega_{osc} \sim \delta m$, oscillations commence.  If the interaction is flavor blind, scatterings are irrelevant, and annihilations proceed as soon as oscillations commence.  This is shown in the upper right panel.  On the other hand, when interactions are flavor sensitive, the decoherence effect of scattering is crucial.  This is shown in the lower two panels.   When there are no scatterings ($\kappa = 0$, left panel), coherence is maintained, but annihilation cannot proceed through flavor sensitive interactions, because $\mbox{det}([{\cal H}_0,Y]) = 0$.  When scatterings are present (right panel), coherence is lost, and annihilations proceed as soon as the scattering rate drops below $H$.  Figure from Ref.~\cite{Tulin:2012re}.}
\label{fig:oscillations}
\end{center}
\end{figure}

Coherence is lost only through scattering via flavor sensitive interactions, as we can see through the relevant term in the Boltzmann equation, which is only non-zero for flavor-sensitive interactions:
\be
\frac{\Gamma_-}{2} \big[ O_-, \big[ O_-,\, {Y} \big] \big] = 2\Gamma_- \left( \begin{array}{cc} 0 & Y_{12} \\ Y_{21} & 0 \end{array} \right).
\label{eq:scattering}
\ee
This term damps $Y_{12}, Y_{21} \to 0$ in Eq.~\ref{Boltzmann} causing decoherence of DM oscillations.  In addition, when this term is large enough, it prevents oscillations from occurring.
The results can be summarized in the series of plots shown in Fig.~\ref{fig:oscillations}, where the effect of coherence and the flavor of the interactions can be seen explicitly.  

\section{Evolution of Astrophysical Objects}
\label{sec:cosmoconstraints}

If the DM retains its cosmological asymmetry late into the universe, unique ``smoking gun'' signatures for ADM may be identified.  In this section we consider the evolution of astrophysical objects.  If ADM retains its asymmetry until today, it may accumulate in large numbers in the center of astrophysical objects.  The amount of ADM that accumulates is proportional to the local density of DM, the scattering cross-section of the DM off the baryons in the object, and the depth of the gravitational potential well of the astrophysical objects.  As we will see, astrophysical objects place substantial constraints on ADM scattering cross-sections on baryons, especially if the DM is a scalar.   We consider the specific cases of constraints from neutron stars, the Sun, and brown and white dwarves. 

\subsection{Neutron Stars}

Besides black holes, neutron stars are the densest astrophysical objects in the universe, and as such provide ideal environments for collecting ADM.  Neutron stars are supported by Fermi degeneracy pressure.  If ADM is bosonic, however, this degeneracy pressure does not exist, so that if the ADM collected in the neutron star exceeds the Chandrasekhar mass, a black hole may form that could eat the neutron star (and the pulsar that envelopes it).  The existence of neutron stars in DM rich regions of the sky can thus constrain regions of scalar ADM parameter space.

While the constraints we focus on below apply only to {\em scalar} ADM,\footnote{The constraints can be weakened or lifted if a repulsive self-interaction is present, as we discuss below.} if an ${\cal O}(1)$ fraction of the neutron star's mass is collected in fermionic ADM  \cite{Goldman:2013qla} or in mirror DM \cite{Sandin:2008db}, then significant modifications of the star can also result.  Very large scattering cross-sections (typically well in excess of the bounds from direct detection experiments) are necessary to accumulate so much DM, however, so that only for very light ($m_X \lesssim 1 \mbox{ GeV}$) fermionic ADM could such conditions be achieved.  On the other hand, a much smaller amount of bosonic ADM accumulated in the neutron star can lead to the formation of a black hole.  Since the constraints are much stronger on bosonic ADM scattering on neutrons, we focus on the scalar ADM case in the remainder of this subsection.

In order to compute the constraint from black hole formation there are four basic stages to the evolution of a neutron star with bosonic ADM:
\begin{itemize}
\item Dark matter collection.  The capture rate in the neutron star is simply proportional to the scattering cross-section off of nucleons, so that evolution of the neutron star in the presence of ADM depends only on the scattering rate and the DM mass.
\item Thermalization.  Once captured, DM rapidly evolves into a thermal ADM sphere, where the size of the thermal sphere is set by the gravitational potential of the neutrons.
\item Self-gravitation and BEC formation.  Enough DM must accumulate so that its self-gravity overwhelms the neutron gravity.  Self-gravitation can occur more quickly if the critical number to form a Bose-Einstein Condensate (BEC) occurs first, so that typically the self gravity condition is met through the formation of a BEC. 
\item Collapse to black hole.  If the self-gravitation condition is met, and the number of DM particles exceeds the Chandrasekhar number, then a black hole forms.  If the rate at which the black hole consumes the neutron star is faster than the rate at which the black hole evaporates, the neutron star will disappear into the black hole.  The existence of the neutron star thus constrains the ADM.
\end{itemize}

The details for each of these stages are as follows.  The subject was first studied in \cite{Goldman:1989nd}, though we follow the discussion from \cite{McDermott:2011jp}, and refer the reader to \cite{Kouvaris:2010vv,Kouvaris:2010jy,Kouvaris:2011fi} for similar calculations. Including the effect of the gravitational field on the BEC transition can improve the constraints that we discuss here by an order of magnitude \cite{Jamison:2013yya}. 

{\bf Collection}: the total number of DM particles collected in the star is a function of the scattering cross-section off nucleons, $\sigma_{X n}$, so that, for $m_X \gtrsim 1 \mbox{ GeV}$ the total number of DM particles in the star is
\beq
N_X \simeq 2.3 \times 10^{45} \left(\frac{10 \mbox{ GeV}}{m_X}\right)\left(\frac{\rho_X}{10^3 \mbox{ GeV/cm}^3}\right)\left(\frac{\sigma_{Xn}}{2 \times 10^{-45} \mbox{ cm}^2}\right)\left(\frac{t}{10^{10} \mbox{ years}}\right),
\eeq
 When the DM mass is less than 1 GeV, the nucleon degeneracy effect becomes important, and we refer the reader to \cite{McDermott:2011jp} for details in that case.
 
{\bf Thermalization}: An estimate of the thermalization time gives \cite{Bertoni:2013bsa}
\beq
t_{th} \simeq 3750 \mbox{ years} \left(\frac{\gamma}{(1+\gamma)^2}\right)\left(\frac{2 \times 10^{-45} \mbox{ cm}^2}{\sigma_n}\right)\left(\frac{10^5 K}{T}\right),
\eeq
where $\gamma = m_X/m_n$.  This is usually rapid in comparison to the lifetime of the neutron star.  

{\bf Self-Gravitation}: Self-gravitation occurs when $\rho_X > \rho_B$, so that the total number is
\beq
N_{self} \simeq 4.8 \times 10^{41} \left(\frac{100 \mbox{ GeV}}{m_X}\right)^{5/2}\left(\frac{T}{10^5 K}\right)^{3/2}.
\eeq
This is to be compared against the Chandrasekhar number:
\beq
N_{Cha}\simeq 1.5 \times 10^{34} \left(\frac{100 \mbox{ GeV}}{m_X}\right)^2
\eeq
For $m_X \lesssim 10^{17} \mbox{ GeV}(T/10^3~K)^3$, we always have $N_{self} \gtrsim N_{Cha}^{boson}$, so that once self-gravitation occurs, collapse to a black hole proceeds.
On the other hand, self-gravitation can happen much faster if a BEC forms first, which occurs if
\beq
N_X \simeq 1.0 \times 10^{36} \left(\frac{T}{10^5~K}\right)^3.
\eeq
Except for heavy DM, the BEC thus occurs first.  The BEC density is much higher than the self-gravity density.  Thus as soon as the number of particles in the ground state exceeds the Chandrasekhar number, black hole formation occurs. 

{\bf Collapse to Black Hole}: In order for the black hole to continue to grow to consume the neutron star, the accretion rate must exceed the Hawking evaporation rate.  The Bondi Hoyle accretion rate is $\pi (G M_{BH}/v_s^2)^2 \rho_B v_s$, with the sound speed $v_s \sim 10^5 \mbox{ km/s}$ where $\rho_B$ the baryon energy density.  The Hawking evaporation rate is $1/(15360 \pi G^2 M_{BH}^2)$.  Balancing these against each other, the critical black hole mass where the black hole will continue to grow is $M_{BH}^{crit} \simeq 1.2 \times 10^{37} \mbox{ GeV}$.   

\begin{figure}
\begin{center}
\includegraphics[width=0.6\textwidth]{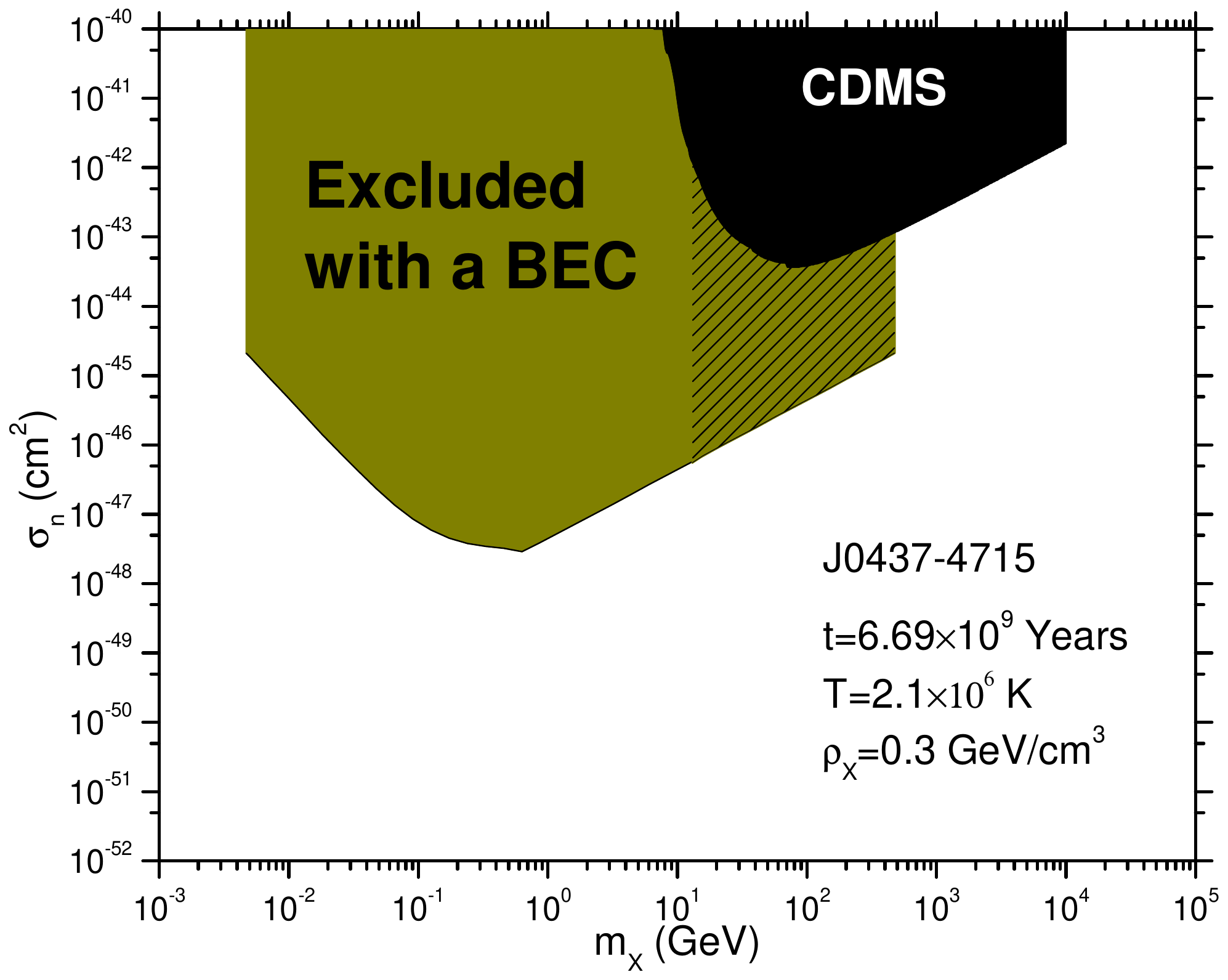}
\caption{Excluded scattering cross-section off a nucleon from Bose-Einstein (and eventual black hole) formation in the neutron star J0437-4715.  In the hatched region constraints are lifted because the mini-black hole evaporates.  For comparison, the constraint from CDMS-II is shown.  From the analysis of \cite{McDermott:2011jp}.}
\label{ns}
\end{center}
\end{figure}

Since the formation of the black hole is ultimately dependent only on the accumulation of bosonic ADM, and hence the DM-nucleon scattering cross-section, the constraints can be mapped in the direct detection plane.  We show the constraints on the scattering cross-section as a function of mass in Fig.~\ref{ns}.  For low mass DM, below where most direct detection experiments are sensitive, neutron stars provide very strong constraints on scalar ADM, though the constraints are lifted due to black hole evaporation over most of the mass range of interest for direct detection experiments.

Self-interactions change this story significantly. For example, a repulsive $\lambda |X|^4/4$ interaction can dramatically change the Chandrasekhar number
\beq
N_{Cha} = \frac{2 M_p^2}{\pi m_X^2}\left( 1 + \frac{\lambda}{32 \pi}\frac{M_p^2}{m_X^2}\right)^{1/2},
\eeq
since the ratio $M_p^2/m_X^2 \sim 10^{36}$ for 1 GeV ADM.  In fact, if $\lambda \gtrsim 10^{-16}$, the neutron star bounds are lifted \cite{Bell:2013xk}.  In addition, $\lambda |X|^4$ type interactions are always generated through any force mediator that could give rise to DM-nucleon scattering that is the source of the constraint.  The {\em sign} of the interaction is, however, important.  A scalar mediator will generate an attractive coupling that will only strengthen the constraint \cite{Guver:2012ba,Bramante:2013hn}.

\subsection{The Sun}

ADM may modify energy transport in the Sun, and hence change neutrino production in the Sun \cite{Spergel:1984re,Griest:1986yu,Taoso:2010tg,Cumberbatch:2010hh,Frandsen:2010yj,Iocco:2012wk,Lopes:2012af}.  We focus on the discussion of \cite{Taoso:2010tg}, and refer the reader to the other papers for further details. The capture rate in the Sun can be approximated by
\begin{eqnarray}
C^\odot & \simeq & 1.3 \times 10^{25} \mbox{s}^{-1}\left(\frac{\rho_{DM}}{0.3\mbox{ GeV/cm}^3}\right)\left(\frac{270 \mbox{ km/s}}{\bar v}\right)\left(\frac{1 \mbox{ GeV}}{m_{DM}}\right)  \\ \nonumber
& \times & \left[\left(\frac{\sigma_H}{10^{-40} \mbox{ cm}^2}\right)S(m_{DM}/m_H) + 1.1 \left(\frac{\sigma_{He}}{16 \times 10^{-40} \mbox{ cm}^2}\right) S(m_{DM}/m_{He})\right],
\end{eqnarray}
where $S \approx 0.9-1.0$ for DM in the 1-10 GeV mass range.\footnote{Heavier elements give an ${\cal O}(1)$ correction to this number which is not important for the purposes of this discussion.}  For ADM with mass below approximately 3 GeV, DM evaporation effects in the Sun lift the constraints we discuss below.

While less dramatic effects can happen at lower DM densities or lower DM capture rates, the most significant effects happen when the DM transport energy $\epsilon_{trans}$ becomes of the same size as the nuclear energy generation term $\epsilon_{nuc}$.  This occurs when the number of accumulated ADM particles exceeds $n_{tot} \sim 5 \times 10^{47}$.  For spin-dependent scattering cross-sections of $\sigma_{SD} \simeq 10^{-37} \mbox{ cm}^2$ off nucleons, this occurs only in main sequence stars with local DM densities close to $10^3 \mbox{ GeV/cm}^3$, which is not achieved locally.  There may, however, still be an effect on main sequence stars in other parts of the galaxy which are more DM rich, so it is still worth considering the impact on main sequence stars. 

The integral of these energy transports over the volume of the star, $L_{trans} \equiv \int |\epsilon_{trans}| dV$, determines the evolution of the star.  When $L_{trans} < L_{nuc}$, the star compensates for the nuclear energy being lost in the core by increasing it outside the core.  When $L_{trans} > L_{nuc}$, this compensation is not possible, so the envelope itself must contract.

\begin{figure}
\begin{center}
\includegraphics[width=0.5\textwidth]{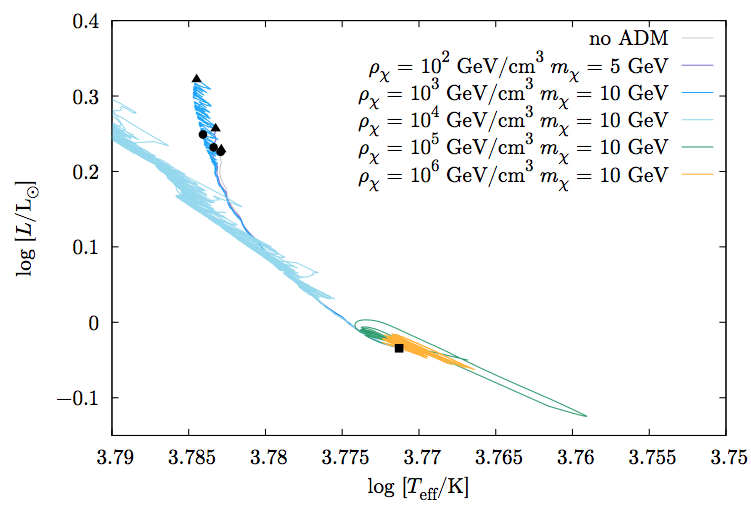}\includegraphics[width=0.5\textwidth]{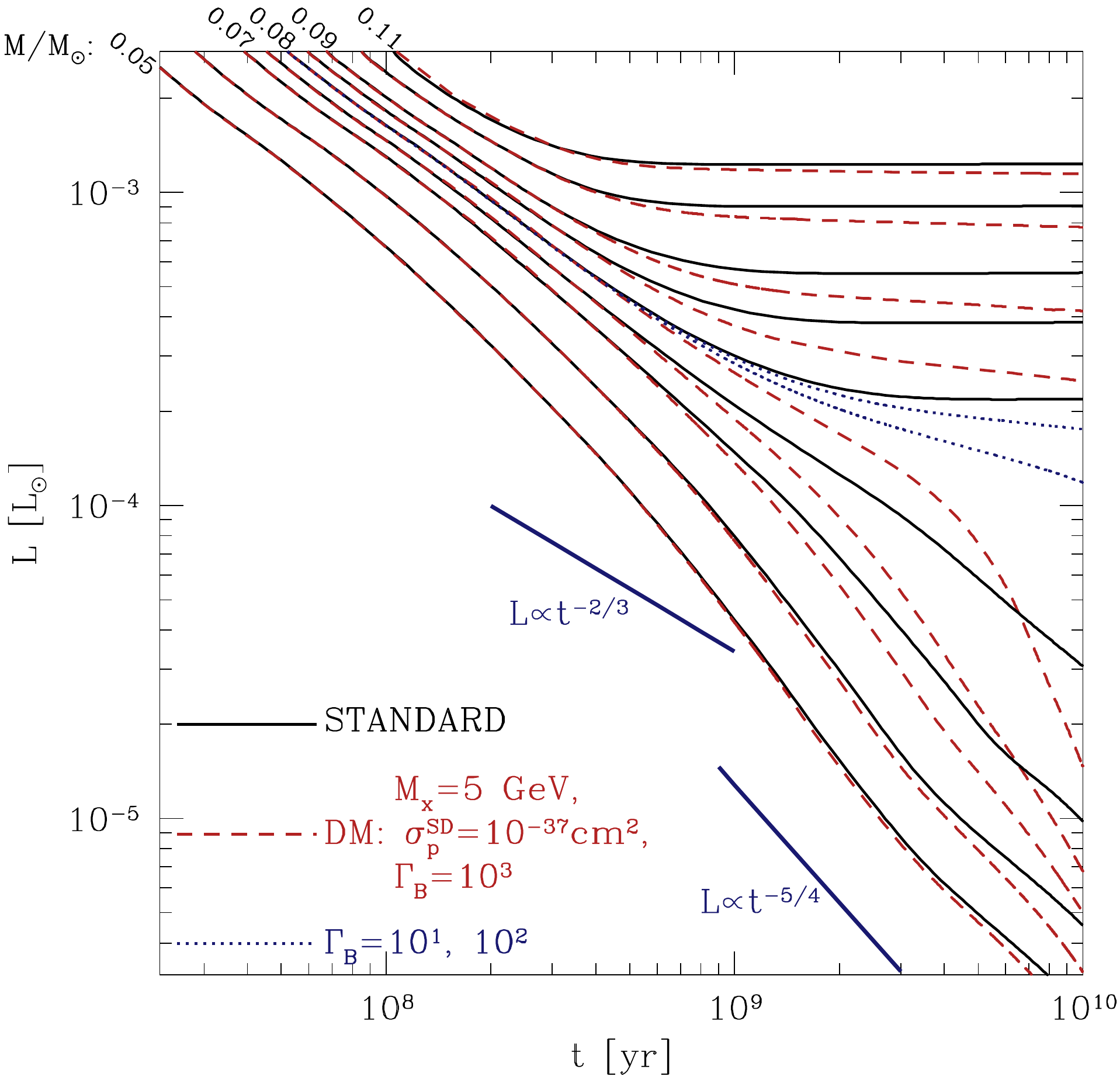}
\caption{{\em left:} The Hertzsprung-Russell diagram for a $1~M_\odot$ star, with $\sigma_{SD} = 10^{-37} \mbox{ cm}^2$.  For a moderate amount of DM accumulation, the star compensates for nuclear energy being lost in the core by increasing it outside the core, making the star hotter and more luminous.  This is shown by the darkest blue curve with $\rho_{DM} = 10^3 \mbox{ GeV/cm}^3$.   Eventually this is no longer possible, and the star contracts and cools, which is shown by the evolution along the green and orange curves.  This yields a dramatic change in the usual evolution of the star that may be observable in a DM dense region.   From \cite{Taoso:2010tg}. {\em right}: Impact of ADM on brown dwarves.  Solid curves show the standard evolution of low mass stars between $0.05$ and $0.11 M_\odot$.  The main sequence is reached as $t$ increases when the luminosity stabilizes to a constant value, indicating that hydrogen burning has been ignited.  In the standard case, this occurs for $M > 0.08~M_\odot$.  By contrast, when DM is added (with boost $\Gamma_B = 10^3$ in comparison to the usual collection rate in the Sun for the same DM parameters, so that enough DM is collected), the red and blue curves result.  Stars between $0.08$ and $0.1 M_\odot$, that entered the main sequence in the standard case, no longer enter the main sequence and instead become brown dwarves.  From \cite{Zentner:2011wx}.}
\label{HRdiag}
\end{center}
\end{figure}

The net effect is that for moderate amounts of captured DM ({\em i.e.}  $L_{trans} < L_{nuc}$ but $\epsilon_{trans} > \epsilon_{nuc}$), the star becomes larger and more luminous.  When $L_{trans} > L_{nuc}$, the envelope finally contracts and becomes cooler, making the star less luminous.  These trends can be seen in the Hertzsprung-Russell diagram in Fig.~\ref{HRdiag}, which shows how stars change their usual evolution to increasingly dramatic modifications as the DM collection increases.

Stars with mass lower than the Sun may also be effective probes of ADM \cite{Casanellas:2012jp}, since the importance of DM energy transport increases as the stellar mass decreases.   Ref.~\cite{Casanellas:2012jp} used $\alpha$ Cen B, via astroseismic measurements, to place meaningful constraints on the DM scattering cross-section on the order of $few \times 10^{-36} \mbox{ cm}^2$ for $\sim 10$ GeV DM, assuming a DM density of 0.4 GeV cm$^{-3}$.  These constraints are stronger than those which can be extracted from the Sun.

\subsection{Brown and White Dwarves}

Similar types of effects can be present in the evolution of low mass stars.  In usual stellar evolution, a low mass star, below $0.08~M_\odot$, evolves to a brown dwarf because the core temperature never becomes hot enough to ignite hydrogen burning.  The accumulation of a significant amount of ADM in the center of a low mass star will affect the energy transport in the star such that, in some cases for stars with masses greater than the usual $0.08~M_\odot$, it cannot reach the main sequence, but instead becomes a brown dwarf \cite{Zentner:2011wx}.  Ref.~\cite{Zentner:2011wx} found that a spin-dependent scattering cross-section of $\sigma_{SD} \sim 10^{-37} \mbox{ cm}^2$ or a spin-independent scattering cross-section of $\sigma_{SI} \sim 10^{-40} \mbox{ cm}^2$, taken with a capture rate boosted by $10^3$ in comparison to that expected in the Sun, will cause a low mass star between $0.08-0.1 M_\odot$ to become a brown dwarf.  This behavior is shown in the right panel of Fig.~\ref{HRdiag}. 

Lastly, accumulation of a large amount of DM in a white dwarf can alter the evolution and cause the star to be much more compact \cite{Leung:2013pra}.  A large amount of DM must be accumulated in the white dwarf in order to  have a substantial effect: in \cite{Leung:2013pra} they found that $10^{-3}~M_\odot$ of 10 GeV DM accumulated in the white dwarf will compactify the core sufficiently that a stable shell model will no longer exist.  Extremely large scattering cross-sections would be necessary to accumulate such a large amount of DM.

\section{Indirect and Direct Detection}
\label{sec:directindirectdetection}

Whether ADM remains symmetric or asymmetric in the universe today, important direct and indirect detection signals are present.  While in many cases signals in direct and indirect detection experiments are not by themselves (in contrast to the effects on the astrophysical objects we discussed in the previous section) indications of an asymmetry in the DM, they are nevertheless important clues to piecing together the nature of the DM.  

We consider three different types of signals in this section.  First, though the DM may have a largely asymmetric abundance in the universe today, a small relic symmetric component can annihilate and give rise to signals in the CMB as well as in indirect detection experiments looking for high energy photons and charged particles.  Second, in some ADM models, the DM may have rare decays to charged particles or neutrinos (in particular through the operators of Eq.~\ref{highdimops}), with lifetime in excess of the age of the universe.  Lastly, in many ADM models, especially those with a light dark force, large scattering cross-sections in direct detection experiments are expected.  We review indirect and direct detection of ADM, beginning with constraints from the CMB.

\subsection{Indirect Detection via the Cosmic Microwave Background}
\label{subsec:CMB}

The number density of DM in the universe today scales inversely with the mass of the DM particle.  Because of the high number density and consequently high annihilation rate of light DM (DM with mass $m_X \lesssim 10 \mbox{ GeV}$), the CMB can be sensitive to light DM annihilation.  The ionizing radiation from DM annihilation can distort the CMB spectrum, so that observations place a constraint on the DM annihilation.  

At the same time, while DM may carry an asymmetry today, a small symmetric component may remain late in the universe.  In the early universe, the annihilation process of $X \bar X$ removes $\bar X$, leaving a prevalence of the asymmetric component $X$.  Depending on the size of the annihilation cross-section, however, a small component of $\bar X$ remains, leading to a small continued annihilation in the universe today.  

The CMB multipole spectrum places a constraint on the annihilation cross-section times ionizing efficiency, $f(z)$, which is $ p_{ann} = f(z) \langle \sigma v \rangle/m_{DM} \lesssim 2.4 \times 10^{-27} \mbox{ cm}^3/\mbox{s}/\mbox{GeV}$ \cite{Finkbeiner:2011dx}.  The ionizing efficiency ranges from around 0.8 for annihilation to $e^+ e^-$, to $\sim 0.3$ for annihilation to most other states (see \cite{Slatyer:2009yq} for a detailed table with ionization efficiencies, and \cite{Galli:2009zc,Galli:2011rz} for updated constraints).  Planck has recently updated this result, though its result is weaker than the WMAP constraint because it does not yet include the full polarization information at intermediate and high multipoles.  They find  $p_{ann} \lesssim 5.5 \times 10^{-27} \mbox{ cm}^3/\mbox{s}/\mbox{GeV}$ \cite{Ade:2013zuv}.  

\begin{figure}
\begin{center}
\includegraphics[width=0.45\textwidth]{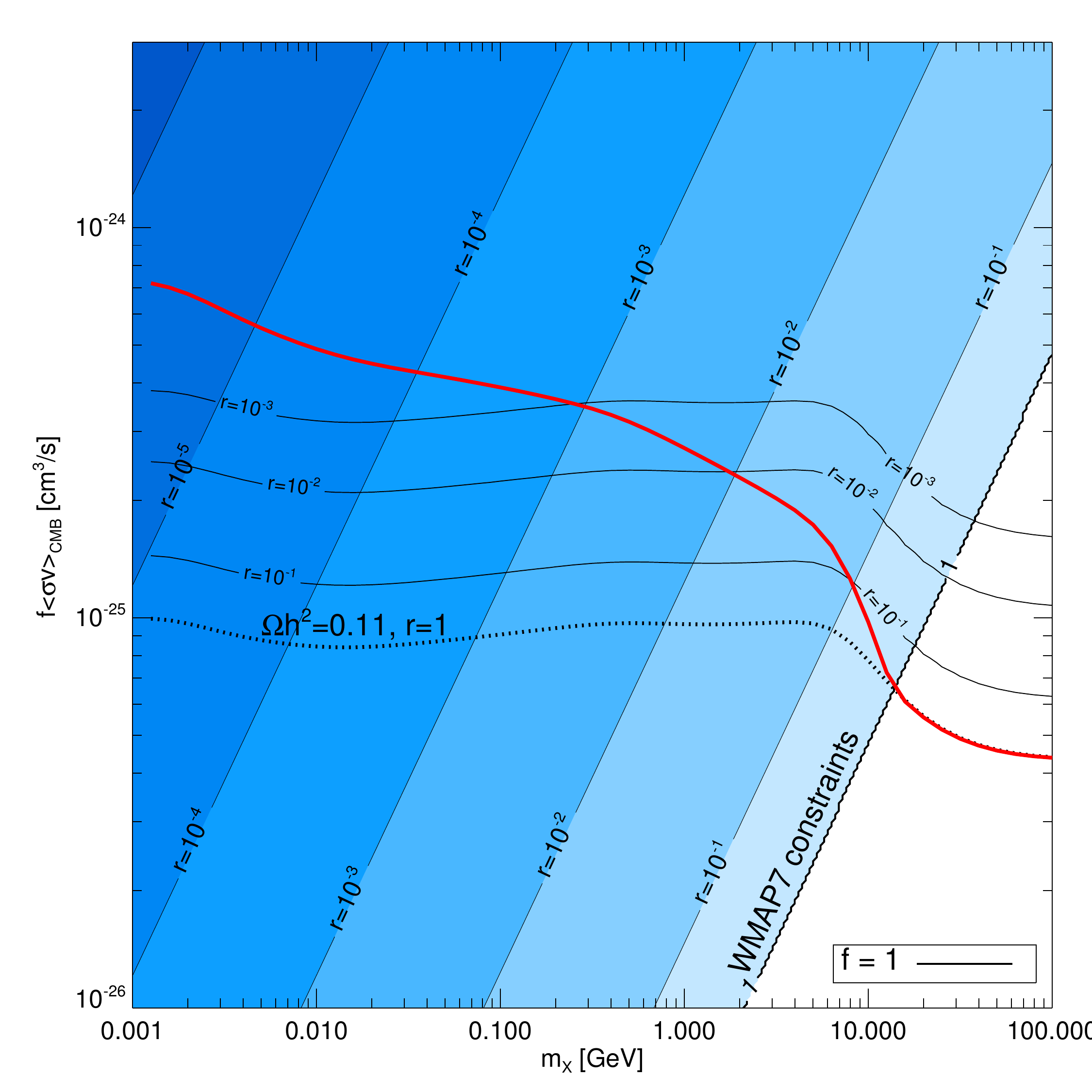}
\caption{The red curve shows the lower bound on the annihilation cross-section times ionizing efficiency of annihilation, $f \langle \sigma v \rangle$, as a function of the DM mass.  The diagonal lines show isocontours of relic symmetric component, $r \equiv \bar X/X$, while the horizontal isocontours show the constraint from the CMB for each choice of $r$.  Where the two isocontours meet is the CMB constraint.  From \cite{Lin:2011gj}.}
\label{CMBconstraints}
\end{center}
\end{figure}

For ADM, the implication is a constraint placing a {\em lower} bound on the annihilation cross-section.  Increasing the annihilation cross-section in the early universe depletes the $\bar X$ component against which $X$ could later annihilate, lifting the CMB constraints.  Assuming $s$-wave annihilation, the relic density of $\bar X$ is computed using the methods of Sec.~\ref{sec:annihilation}.  The annihilation rate of $X-\bar X$ is then computed, and a constraint derived depending on the ionization efficiency.  The results are shown in Fig.~\ref{CMBconstraints}.  Depending on the mass of the DM, the CMB places the constraint $f \langle \sigma v \rangle \gtrsim 10^{-25} \mbox{ cm}^3/\mbox{s}$, for DM that retains its asymmetry until today.  The results in this figure can be easily scaled as new experimental results are obtained, taking into account that the constraint in the figure corresponds to a bound $p_{ann} \lesssim 2.4 \times 10^{-27} \mbox{ cm}^3/\mbox{s}/\mbox{GeV}$.

\subsection{Indirect Detection in the Milky Way}

While, naively, asymmetric DM is not self-annihilating, as we discussed in Sec.~\ref{sec:oscillations}, its asymmetry can be washed out via oscillations after the relic density has already been set by its primordial chemical potential.  Thus annihilations may recouple today, and give rise to an annihilation rate enhanced relative to that expected from a thermal relic, as in the models of Refs.~\cite{Cohen:2009fz,Cai:2009ia}.  However, even when the asymmetry persists until today,  as we reviewed in Sec.~\ref{subsec:CMB} for the case of $s$-wave annihilation, the small relic symmetric $X$ component implies a {\em lower} bound on the annihilation cross-section from the CMB.  If we put aside the constraint from the CMB, the presence of a significant symmetric (sub-)component of ADM could thus give rise to a {\em larger} annihilation rate from ADM than would be expected from thermal DM \cite{Gelmini:2013awa}, even when the DM density is dominated by the asymmetric component.  

A variety of other signals can arise.  For example, for the higher dimension operator models of Sec.~\ref{subsec:highdimops}, four fermion operators arise, and the DM may be long lived if the scale of the operator $M$ is high.  For dimension six four fermion operators from Eq.~\ref{highdimops}, the lifetime is approximately $10^{26} \mbox{ s}$ for 10 GeV DM and $M=10^{14} \mbox{ GeV}$.  When the DM decays, photons may be radiated off the final state particles. 
It is also possible that DM may scatter off of charged cosmic rays and give rise to energetic photons \cite{Profumo:2011jt}, though the cross-sections need to be very large to produce a detectable signal.

ADM may also leave imprints in charged cosmic rays.  For example, it has been suggested \cite{Chang:2011xn,Masina:2011hu,Masina:2013yea}, that ADM decays could give rise to the charge asymmetry observed by PAMELA, Fermi, and AMS-02, through the asymmetric decays such as $X \rightarrow \chi^- \mu^+$ or $X \rightarrow H^- \ell^+$ with $H^- \rightarrow \tau^- \bar \nu$.  Asymmetric decays through the higher dimension operator $LL E^c$ were also considered in \cite{Feng:2013vva}.  Self-interacting ADM may emit force carriers in scattering processes \cite{Pearce:2013ola}, which subsequently decay, producing charged particles and photons, or creating a ``dark disk'' \cite{Fan:2013yva,Fan:2013tia}.

Lastly, high energy neutrinos could be a signal for ADM, and \cite{Feldstein:2010xe} showed that the scale of dimension six ADM operators (fermionic forms of the operators in Eq.~\ref{highdimops}), could be bounded to be $M \gtrsim 10^{12}-10^{13} \mbox{ GeV}$, which is modestly weaker than the bounds from photons.

\subsection{Direct Detection}

The ADM mechanism itself, through the operators Eq.~\ref{highdimops}, need not provide a direct detection (DD) signal, though in some cases, depending on the UV completion, it will.  For example, the operator $X u^c d^c d^c$ has, as one possible UV completion, interactions of the form $\lambda_X X u^c U + \lambda_U U^c d^c d^c + m_U U U^c$.  The heavy particle $U$ may mediate $t$-channel scattering off of nucleons, though the scattering cross-section is generically very small, even if the state is near the weak scale \cite{Gelmini:1986zz}:
\begin{eqnarray}
\sigma_{n,X}&  \simeq & \frac{\lambda_X^4 \mu_n^2}{\pi m_U^4} \frac{\left[\frac{2}{3} Z^2 + \frac{1}{3}(A-Z)^2 \right]}{\left[Z^2 + (A-Z)^2 \right]} \nonumber \\ 
 &\simeq & 5 \times 10^{-45} \mbox{ cm}^2\left(\frac{\lambda_X^2/m_U}{100 \mbox{ TeV}}\right)^2 \left(\frac{1 \mbox{ TeV}}{m_U}\right)^2, 
 \label{tchannelscattering}
\end{eqnarray}
where $\mu_n$ is the DM-nucleon reduced mass, we have taken $A=28,~Z=14,~m_X = 10$ GeV, and we have inserted meson oscillation constraints on $\lambda^2/m_U$ of around 100 TeV. 

Models with dark forces, as discussed in Sec.~\ref{sec:hidden sectors}, may also provide a direct detection mediator.  In these models, the dark force often mixes with hypercharge through a term $\epsilon F'_{\mu \nu} F^{\mu \nu}$.   This mixing both sets the mass scale of the DM, as in Sec.~\ref{sec:hidden sectors}, and provides a connection to the visible sector for direct detection.  The scattering cross-section via kinetic mixing off the proton is
\begin{eqnarray}
\sigma_{p,X} & = &  \frac{4}{\pi}\frac{g_Y^2 g_X^2 c_W^4 \epsilon^2 \mu_n^2}{m_{\gamma_d}^4} \\ \nonumber
& \simeq & 4 \times 10^{-41} \mbox{ cm}^2 \left(\frac{g_X}{0.1}\right)^2  \left(\frac{\epsilon}{10^{-3}}\right)^2\left(\frac{10 \mbox{ GeV}}{m_{\gamma_d}}\right)^4, 
\end{eqnarray}
where $g_X$ is the $U(1)_X$ gauge coupling, $\mu_n$ is the DM-nucleon reduced mass, and $m_{\gamma_d}$ is the mass of the dark photon.  This is naturally in the same range observed by DAMA \cite{Bernabei:2010mq}, CoGeNT \cite{Aalseth:2010vx,Aalseth:2011wp,Aalseth:2012if} and CDMS \cite{Agnese:2013rvf}.  This was utilized in several models for these signals \cite{Zurek:2008qg,Essig:2010ye,Cohen:2010kn}.

The interest in ADM has been partly driven by the hints for DM from these experiments, which prefer a DM mass in the 7-10 GeV window and a scattering cross-section around $10^{-40}-10^{-41} \mbox{ cm}^2$.  The apparent tension between the possible signals and the null results from other experiments, notably XENON10 \cite{Angle:2011th} and XENON100 \cite{Aprile:2012nq}, has driven efforts to reduce the tension by scattering through operators with non-standard momentum dependence \cite{Chang:2009yt,Fitzpatrick:2010br}, or by tuning away the coupling to xenon by isospin violation \cite{Feng:2011vu,Okada:2013cba}.  In general, ADM has been cited as motivation for a wide range of studies on the CoGeNT, DAMA and CDMS signals \cite{Graham:2012su,Hooper:2012ft,Kelso:2011gd,Essig:2011nj,Krohn:2010ad}.  The class of mirror DM models has also been utilized as an explanation, {\em e.g.} \cite{Foot:2003iv,Foot:2005ic,Foot:2008nw,Foot:2011pi}.

A further explicit connection may be made between the force carrier to which the DM annihilates, Eq.~\ref{annihilateThermalAbundance}, to remove its symmetric relic abundance, and scattering through this same force carrier \cite{Lin:2011gj}.  Sufficient annihilation fixes a lower bound on the coupling, $\alpha_X \equiv g_X^2/(4 \pi) \gtrsim 5.2 \times 10^{-5} (m_X/\mbox{ GeV})$, between the force mediator and the DM.  The total scattering cross-section depends on the product of $\alpha_X$ with the coupling to nucleons, $g_n$: 
\begin{equation}
\sigma_{n,X} = 4 \alpha_X g_n^2 \frac{\mu_n^2}{m_\phi^4}.
\end{equation}
A lower bound on $g_n$ can obtained by requiring thermal equilibrium at some point in the early universe between the visible and hidden sectors.  One finds that this leads to a lower bound on the scattering cross-section \cite{Lin:2011gj}
\beq
\sigma_{n,X} \gtrsim 10^{-48} \mbox{ cm}^2 \left(\frac{m_X}{\mbox{ GeV}}\right)^4 \left(\frac{\mbox{GeV}}{m_\phi}\right)^6\left(\frac{\mu_n}{0.5 \mbox{ GeV}}\right)^2.
\eeq
Thus direct detection experiments could rule out an ADM sector thermalized with the visible sector through a light hidden gauge boson.

\section{Collider Signatures and Constraints}
\label{sec:colliderconstraints}

\subsection{Monojets and Extended Supersymmetric Cascade Decays}

ADM leads to a potentially rich array of DM signals at high energy colliders.  Though the DM mass lies well below the weak scale, it may be difficult to produce the DM at the collider because it may be weakly coupled to the SM, as depicted in Fig.~\ref{ADMSchematic}, which takes its structure from ``Hidden Valley'' models \cite{Strassler:2006im,Strassler:2006ri}.  The $x$-axis schematically represents inaccessibility and the $y$-axis represents energy.  The SM is highly accessible and its ground state resides at low energy.  The ADM sector also has a ground state particle that resides at low energy, but as is evident from the difficulty of directly detecting the DM sector, it is highly inaccessible to us.  Between the SM and DM sector is a barrier.  This barrier may either be massive states (mass at the electroweak scale or above) that couple to both the SM and to the ADM sector, or light states that couple weakly to the SM sector so that they escape constraints from lower energy machines.  Once passing over the barrier into the DM sector, an immense amount of structure may be present.  We already discussed in Sec.~\ref{sec:hidden sectors} the importance of dark forces, and the potential need for a dark Higgs mechanism to give mass to the DM and set the scale for the DM sector, and we discussed the impacts on structure formation.  Here we elaborate on some of the implications of this picture for collider searches.

Perhaps the simplest direct way to constrain DM at a collider is through initial state radiation plus missing energy \cite{Petriello:2008pu,Gershtein:2008bf}, where the missing energy (MET) signifies DM production.  Monojet plus MET \cite{Beltran:2010ww,Goodman:2010yf,Goodman:2010ku,Bai:2010hh,Fox:2011pm} searches have become a powerful tool in the search for DM.  It is important to recognize the limit of the effectiveness of these searches, however.  Their constraints become very weak if the mediator of the scattering is lighter than the weak scale, as shown explicitly in \cite{An:2012va}.  Thus direct detection and monojet searches are complementary probes of DM.

ADM may also be made supersymmetric, as an extension of the MSSM.  In this case, the lightest particle in the MSSM, the lightest ordinary supersymmetric particle (LOSP), is unstable to decay into the ADM sector.  In the simplest ADM models, this means direct decay into the DM particle plus additional SM particles.  For example, in one of the models of \cite{Kaplan:2009ag}, reviewed in Sec.~\ref{subsec:highdimops}, the operator $W = X^2 L H/M$ had two DM particles, one stabilized by $R$-parity and one stabilized by the additional $Z_2$.  Supposing that the neutralino is the LOSP, the LOSP decays to $\bar{\nu} \tilde{X} \tilde{X}$, and the lifetime of this process is {\em approximately} \cite{Kaplan:2009ag},
\beq
c \tau(\chi^0 \rightarrow \bar \nu \tilde X \tilde X) \sim \mbox{cm} \left(\frac{M}{10^8 \mbox{ GeV}}\right)^2 \left(\frac{m_{\chi^0}}{100 \mbox{ GeV}}\right)^{-3},
\label{neutralinoLH}
\eeq 
which may be collider stable or displaced.  The $X$ scalars subsequently decay via $\tilde X \rightarrow \bar X \nu$.  This decay is entirely visible, so that the true nature of the DM may be mistaken for a neutralino.  

On the other hand, the LOSP may have a different nature, or the operator transferring the asymmetry may be different, giving rise to visible decays.  For example, in the $X^2 L H$ interaction, a $\tilde \tau$ LOSP would decay to $\tilde{\tau}_R \rightarrow \tau \nu \bar X \bar X$ via a virtual neutralino with a lifetime
\begin{equation}
c \tau(\tilde{\tau}_R \rightarrow \tau \nu \bar X \bar X) \sim \mbox{mm} \left(\frac{M}{10^6 \mbox{ GeV}}\right)^2\left(\frac{m}{200 \mbox{ GeV}}\right)^6\left(\frac{m_{\tilde{\tau}}}{100 \mbox{ GeV}}\right)^{-7},
\label{stauLH}
\end{equation}
where a common mass scale $m \sim m_{\tilde{\nu}}\sim m_{\chi^0}$ is assumed.

Replacing the $L H$ transfer interaction with $u^c d^c d^c$ gives rise to  generically collider displaced vertices. For example the neutralino decay through $u^c d^c d^c$ gives rise to an approximate lifetime
\beq
c \tau(\chi^0 \rightarrow q q q \tilde{X} \tilde X) \sim 0.3 \mbox{ mm} \left(\frac{M}{\mbox{TeV}}\right)^4\left(\frac{m}{500 \mbox{ GeV}}\right)^4\left(\frac{m_{\chi^0}}{100 \mbox{ GeV}}\right)^{-9}.
\label{neutralinoudd}
\eeq

\begin{figure}
\begin{center}
\includegraphics[width=0.6\textwidth]{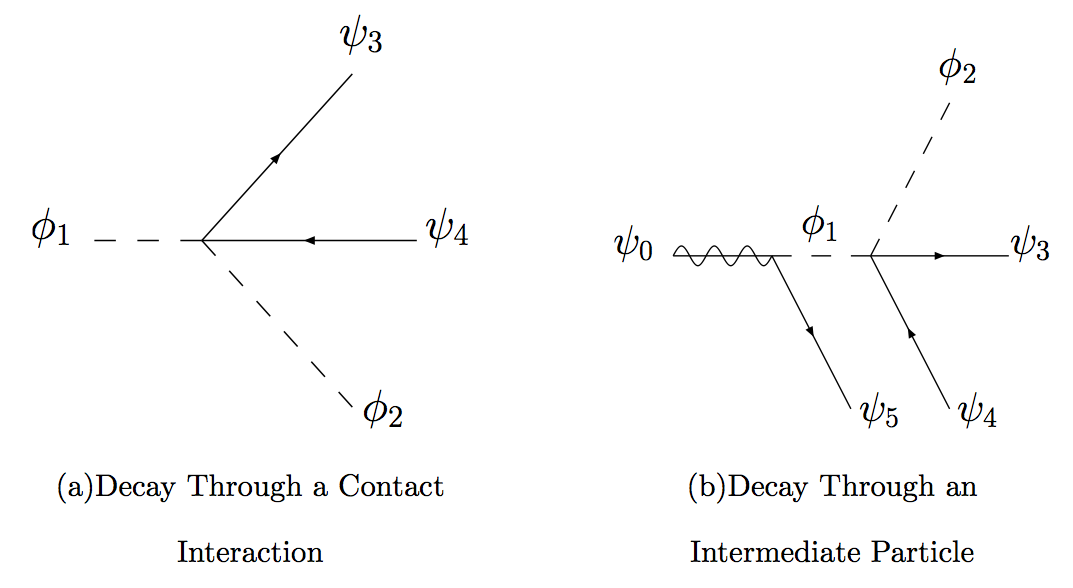}
\caption{Diagrams for the decay of the lightest ordinary supersymmetric particle (LOSP) to the scalar $\tilde X$ and two SM fermions, if the LOSP is a squark ({\em left}) or neutralino ({\em right}).}
\label{ADMDecay}
\end{center}
\end{figure}

$R$-parity alone may be used to stabilize the DM.  In this case $X {\cal O}_{MSSM}$ ({\em e.g.} $W = X u^c d^c d^c/M$) may instead be the transfer operator.  In this case, decay is to a single scalar $\tilde{X}$, through the diagrams in Fig.~\ref{ADMDecay}.  The left-hand diagram dominates when the LOSP is one of the fields in the ${\cal O}_{SM}$ operator, while the right-hand diagram will dominate when the LOSP is a gaugino.  

As the LOSP decays into the DM, the extended supersymmetric cascade decay chains imply a higher multiplicity of visible final state particles than for the same MSSM spectrum.  At the same time, missing energy is reduced, in a similar way as models of R-parity violating, though in ADM, at least one light DM particle results at the bottom of the chain.  Thus, signatures for these models are in between the usual MSSM signatures with jets, leptons and missing energy, and RPV signatures, where the LOSP decays to jets and leptons plus no missing energy.  The collider constraints on MSSM superpartners in ADM models have been shown to be weaker than with no ADM sector \cite{Kim:2013ivd}.  New searches for supersymmetry should be implemented at the LHC to maximize sensitivity.  

\subsection{Flavor Constraints}

For models of ADM mediated through the higher dimension operators reviewed in Sec.~\ref{subsec:highdimops}, there may be important implications for flavor physics, depending on the scale, $M$, of the UV completion of the operators.  As we saw in Eqs.~\ref{neutralinoLH}-\ref{neutralinoudd}, the scale $M$ also correlates with whether the LOSP has prompt decays at the LHC.  If the scale is too high, the LOSP will be collider stable, even if it is not the DM.  We examine for a moment $ {\cal O}_{B-L} = q \ell d^c$.  The UV completion of this operator may include terms such as the following:
\beq
W_D = \lambda_{XD}^i X d_i^c {\cal D} + \lambda_D^{i j} {\cal D}^c q_i \ell_j + M_D {\cal D D}^c,
\eeq
where $i,~j$ are flavor indices, and $\cal D$ is a new heavy state being integrated out to generate the operator in Eq.~\ref{highdimops}.  Depending on the flavor structure of $\lambda_{XD}^i$ and $\lambda_D^{i j}$, this can give rise to meson mixing, $b \rightarrow s \ell^+ \ell^-$, $\mu \rightarrow e \gamma$, $\mu^- \rightarrow e^- e^+ e^-$, $\mu-e$ conversion.  In the absence of a flavor structure, these constraints typically can place a limit on $M_D/\lambda^2$, where $\lambda^2$ is a diagram dependent combination of $\lambda_{XD}^i$ and $\lambda_D^{i j}$.  The constraints remain to be explored in detail in the literature. 

\section{Summary and Outlook}

In the last several years, Asymmetric Dark Matter has become a flourishing subfield of DM research.  This is driven in part by a purely theoretical motivation to explore well-motivated models outside the standard WIMP parameter space, and in part by anomalies in Direct and Indirect Detection.  In Secs.~\ref{sec:models} and \ref{sec:baryo}, we focused on outlining the general mechanisms for transferring or generating an asymmetry in the DM sector, and only relied on specific models for illustrating features of these mechanisms.  As we saw in Sec.~\ref{sec:annihilation} and \ref{sec:hidden sectors}, ADM models in many cases also go hand in hand with hidden sector DM models, where additional structure, in the form of dark gauge and global symmetries, give rise to a wide range of impacts on DM dynamics.  

One of the most important implications of ADM is phenomenological, since theories of DM inform the types of experiments that will be designed and utilized for hunting DM.  Neutron stars, brown dwarves and the Sun constrain properties of ADM, as we reviewed in Sec.~\ref{sec:cosmoconstraints}, but only when the ADM is either a scalar or has a very large scattering cross-section with nucleons.  In addition, the primordial effects of a cosmological asymmetry in the DM symmetry can be washed away through DM-anti-DM oscillations at late times, as we reviewed in Sec.~\ref{sec:oscillations}.  Thus signals in indirect detection may be restored.  Lastly, the signals from the CoGeNT and CDMS experiments in the 7-10 GeV mass window increase the interest in ADM theories, where the scattering rate observed is predicted by scattering through a light dark force, as in Sec.~\ref{sec:directindirectdetection}.  On the other hand, ADM models do not {\em necessarily} predict a scattering cross-section that is observable in direct detection experiments.  Lastly, the presence of ADM as part of supersymmetric models changes collider searches for these states.

ADM has many possible implications for signatures in a wide range of experiments from direct and indirect detection to the LHC.  Once DM is discovered, it is just the beginning of the path to uncovering the potentially rich story of the nature of the dark side of the universe.  Many of the tools of this review will be crucial along that path.

\section*{Acknowledgments}

We thank Moira Gresham, David Morrissey, Michele Papucci and Haibo Yu for careful readings of the draft of this manuscript.  KZ is supported by the DoE under contract de-sc0007859, by NSF CAREER award PHY 1049896, and by NASA Astrophysics Theory grant NNX11AI17G. 

\appendix

\section{Distribution of baryon, lepton and dark matter asymmetries}
\label{subsec:distrib}

A classic calculation, following \cite{Harvey:1990qw}, allows one to accurately compute the relation of the baryon, lepton and dark matter asymmetries via the transfer operators and the electroweak sphalerons which distribute $B$ and $L$ in relation to the conserved $B-L$.  The details of the results depend on the details of the phase transition ({\em e.g.} whether the sphalerons decouple before or after the electroweak phase transition), as well as what fields are in thermal equilibrium when the sphalerons decouple (notably, the top quark).  We review the standard calculation, and give a few results for the new calculation in ADM.

Chemical equilibrium gives rise to relations between the chemical potentials.  The equations for the SM are
\begin{eqnarray}
\mu_W & = & \mu_- + \mu_0~~(W^- \leftrightarrow \phi^- + \phi^0) \\
\mu_{d_L} & = & \mu_{u_L} + \mu_W~~(W^- \leftrightarrow \bar u_L + d_L) \\
\mu_{iL}& =& \mu_i + \mu_W~~(W^- \leftrightarrow \bar\nu_{iL} + e_{iL}) \\
\mu_{uR} & = & \mu_0 + \mu_{uL}~~(\phi^0 \leftrightarrow \bar u_L + u_R) \\
\mu_{dR} & = & -\mu_0 +\mu_W+ \mu_{uL}~~(\phi^0 \leftrightarrow \bar d_L + d_R) \\
\mu_{iR} & = & -\mu_0+\mu_W+\mu_i~~(\phi^0 \leftrightarrow e_{iL} + \bar e_{iR}),
\label{chemeqb}
\end{eqnarray}
where $\phi$ is the Higgs boson.
In the standard calculation, the sphalerons impose the relation
\beq
N_g(\mu_{uL} + 2 \mu_{dL}) + \mu = 0,
\label{sphaleron}
\eeq
where $N_g$ is the number of SM generations and $\mu \equiv \sum_{i=1,N_g} \mu_{i}$.
In addition, the total charge
\beq
Q = 2 N_g(\mu_{u_L} + \mu_{u_R})-N_g(\mu_{d_L}+\mu_{d_R})-\sum_i(\mu_{i L}+\mu_{iR})-4 \mu_W - 2 m \mu_-
\label{Q}
\eeq
must be zero at all times, and the $SU(2)$ charge
\begin{eqnarray}
Q_3  & = & \frac{3 N_g}{2}(\mu_{u_L} - \mu_{d_L}) + \frac{1}{2}\sum_i (\mu_i-\mu_{iL}) - 4 \mu_W - m(\mu_0 + \mu_-) \\ \nonumber
& = & -(2 N_g + m + 4) \mu_W
\label{Q3}
\end{eqnarray}
must be zero when $SU(2)$ is unbroken above the electroweak phase transition, where $m$ is the number of Higgs doublets.  This immediately implies $\mu_W = 0$ above the electroweak phase transition.\footnote{The non-zero chemical potential of the Higgses may be transferred to Higgsinos in supersymmetric models, allowing for asymmetric Higgsino DM \cite{Blum:2012nf}.  This asymmetry is typically too small, however, to account for the observed DM density.}  Below the electroweak phase transition, the condition that $Q_3$ be zero is replaced by the requirement that $\mu_0 = 0$ once the Higgs obtains a vacuum expectation value.  This allows one to write $B$ and $L$ in terms of the conserved quantity $B-L$.

Now this calculation is modified in various ways depending on the nature of the ADM sector, and the presence of a messenger sector near the electroweak scale.  We discuss these modifications here, assuming the DM is charged under a new global symmetry $U(1)_X$.  First, if states charged under $U(1)_X$ are also charged under $SU(2)$, then it is no longer $B+L$ that is violated but rather $B+L + N_X/N_g X$, where $N_X$ is the number of dark doublets charged under $SU(2)$, and dark number $D$ is defined by $X \equiv \mu_X - \mu_{\bar X}$.  Thus the new sphaleron equation is
\beq
N_g(\mu_{uL} + 2 \mu_{dL}) + \mu + N_X \mu_X = 0.
\label{sphaleronmod}
\eeq  

Second, if the asymmetry is instead communicated through higher dimension operators, a new constraint equation of the dark chemical potential is enforced.  For example, through the $X u^c d^c d^c$ model, the relation is
\beq
\mu_X - \mu_{u_R} - 2 \mu_{d_R} = 0,
\label{highchempot}
\eeq  
if only fermions are involved in the interaction.  Supersymmetric interactions enforce relations of the type
\beq
\mu_X^t - \mu_{u_R}^t - 2 \mu_{d_R} = 0,
\eeq  
where a superscript $t$ denotes tilde and indicates the (scalar) superpartner.

Superpartners will also contribute to the total $B$, $L$ or $D$ number if they are in thermal equilibrium at the temperature at which one is evaluating the relations between the chemical potentials.  Note however that a superpartner has twice as many degrees of freedom in the equation.  For example, with all particles (including the top) and super partners in thermal equilibrium, we have for the total baryon number 
\beq
B = N_g (\mu_{u_L}+2 \mu_{u_L}^t) + N_g (\mu_{u_R}+2 \mu_{u_R}^t) + N_g (\mu_{d_L}+2 \mu_{d_L}^t) + N_g (\mu_{d_R} + 2 \mu_{d_R}^t).
\eeq
Similar relations can be written for $L$ and $D$.

Lastly, as particles drop out of thermal equilibrium, they no longer contribute to $B$, $L$ or $D$, though the chemical equilibrium Eqs.~\ref{chemeqb},~\ref{highchempot} and sphaleron Eqs.~\ref{sphaleron},~\ref{sphaleronmod} remain the same.  For example, at temperatures below the top mass, the baryon number is
\beq
B = (N_g-1) \mu_{u_L} + (N_g-1) \mu_{u_R} + N_g \mu_{d_L} + N_g \mu_{d_R}.
\eeq
The equations for $Q$, $Q_3$ (Eqs.~\ref{Q},~\ref{Q3}), $B$ and $D$ are also accordingly modified as particles carrying these charges become exponentially suppressed in number density.

These relations can be utilized rather generically to extract a relation between the DM and proton mass, on a case by case basis, though we gave two examples in Sec.~\ref{subsec:sphalerons}.  In some cases the prediction for the DM mass may be very constrained simply by the conserved quantum numbers of the system \cite{Ibe:2011hq}.   

\bibliography{refs}

\end{document}